\documentclass[twocolumn]{aastex631}
\usepackage{CJKulem}
\shortauthors{Ma et al.}
\graphicspath{{./}{figures/}}

\begin{document}

\title{A detailed view of low-frequency quasi-periodic oscillation in the broadband 0.2--200\,keV with \emph{Insight}-HXMT and NICER}

\author[0000-0002-2032-2440]{X. Ma}
\affiliation{Key Laboratory of Particle Astrophysics, Institute of High Energy Physics, Chinese Academy of Sciences, Beijing 100049, People’s Republic of China}
\correspondingauthor{L. Tao}
\email{taolian@ihep.ac.cn}
\correspondingauthor{S.N. Zhang}
\email{zhangsn@ihep.ac.cn}
\correspondingauthor{L. Zhang}
\email{zhangliang@ihep.ac.cn} 
\affiliation{University of Chinese Academy of Sciences, Chinese Academy of Sciences, Beijing 100049, People’s Republic of China}
\author[0000-0003-4498-9925]{L. Zhang}
\affiliation{Key Laboratory of Particle Astrophysics, Institute of High Energy Physics, Chinese Academy of Sciences, Beijing 100049, People’s Republic of China}
\author[0000-0002-2705-4338]{L. Tao}
\affiliation{Key Laboratory of Particle Astrophysics, Institute of High Energy Physics, Chinese Academy of Sciences, Beijing 100049, People’s Republic of China}
\author[0000-0001-5238-3988]{Q.C. Bu}
\affiliation{Institut für Astronomie und Astrophysik, Kepler Center for Astro and Particle Physics, Eberhard Karls Universität, Sand 1, D-72076 Tübingen, Germany}
\author[0000-0002-9796-2585]{J. L. Qu}
\affiliation{Key Laboratory of Particle Astrophysics, Institute of High Energy Physics, Chinese Academy of Sciences, Beijing 100049, People’s Republic of China}
\affiliation{University of Chinese Academy of Sciences, Chinese Academy of Sciences, Beijing 100049, People’s Republic of China}
\author[0000-0001-5586-1017]{S.N. Zhang}
\affiliation{Key Laboratory of Particle Astrophysics, Institute of High Energy Physics, Chinese Academy of Sciences, Beijing 100049, People’s Republic of China}
\affiliation{University of Chinese Academy of Sciences, Chinese Academy of Sciences, Beijing 100049, People’s Republic of China}
\author{D.K., Zhou}
\affiliation{Key Laboratory of Particle Astrophysics, Institute of High Energy Physics, Chinese Academy of Sciences, Beijing 100049, People’s Republic of China}
\affiliation{University of Chinese Academy of Sciences, Chinese Academy of Sciences, Beijing 100049, People’s Republic of China}
\author[0000-0002-3515-9500]{Y. Huang}
\affiliation{Key Laboratory of Particle Astrophysics, Institute of High Energy Physics, Chinese Academy of Sciences, Beijing 100049, People’s Republic of China}
\author[0000-0002-5203-8321]{S.M. Jia}
\affiliation{Key Laboratory of Particle Astrophysics, Institute of High Energy Physics, Chinese Academy of Sciences, Beijing 100049, People’s Republic of China}
\affiliation{University of Chinese Academy of Sciences, Chinese Academy of Sciences, Beijing 100049, People’s Republic of China}
\author[0000-0003-0274-3396]{L.M. Song}
\affiliation{Key Laboratory of Particle Astrophysics, Institute of High Energy Physics, Chinese Academy of Sciences, Beijing 100049, People’s Republic of China}
\affiliation{University of Chinese Academy of Sciences, Chinese Academy of Sciences, Beijing 100049, People’s Republic of China}
\author{S. Zhang}
\affiliation{Key Laboratory of Particle Astrophysics, Institute of High Energy Physics, Chinese Academy of Sciences, Beijing 100049, People’s Republic of China}
\affiliation{University of Chinese Academy of Sciences, Chinese Academy of Sciences, Beijing 100049, People’s Republic of China}
\author[0000-0002-2749-6638]{M.Y. Ge}
\affiliation{Key Laboratory of Particle Astrophysics, Institute of High Energy Physics, Chinese Academy of Sciences, Beijing 100049, People’s Republic of China}
\affiliation{University of Chinese Academy of Sciences, Chinese Academy of Sciences, Beijing 100049, People’s Republic of China}
\author{H.X. Liu}
\affiliation{Key Laboratory of Particle Astrophysics, Institute of High Energy Physics, Chinese Academy of Sciences, Beijing 100049, People’s Republic of China}
\affiliation{University of Chinese Academy of Sciences, Chinese Academy of Sciences, Beijing 100049, People’s Republic of China}
\author{Z.X. Yang}
\affiliation{Key Laboratory of Particle Astrophysics, Institute of High Energy Physics, Chinese Academy of Sciences, Beijing 100049, People’s Republic of China}
\affiliation{University of Chinese Academy of Sciences, Chinese Academy of Sciences, Beijing 100049, People’s Republic of China}
\author{W. Yu}
\affiliation{Key Laboratory of Particle Astrophysics, Institute of High Energy Physics, Chinese Academy of Sciences, Beijing 100049, People’s Republic of China}
\affiliation{University of Chinese Academy of Sciences, Chinese Academy of Sciences, Beijing 100049, People’s Republic of China}
\author[0000-0002-8442-9458]{E. S. Yorgancioglu}
\affiliation{Institut für Astronomie und Astrophysik, Kepler Center for Astro and Particle Physics, Eberhard Karls Universität, Sand 1, D-72076 Tübingen, Germany}









\begin{abstract}

We report the X-ray timing results of the black hole candidate MAXI J1820+070 during its 2018 outburst using the Hard X-ray Modulation Telescope (\emph{Insight}-HXMT) and Neutron Star Interior Composition Explorer Mission (NICER) observations. Low frequency quasi-periodic oscillations (LFQPOs) are detected in the low/hard state and the hard intermediate state, which lasted for $\sim$90 days. Thanks to the large effective area of \emph{Insight}-HXMT at high energies and NICER at low energies, we are able to present the energy dependence of the LFQPO characteristics and phase lags from 0.2\,keV to 200\,keV, which has never been explored by previous missions. We find that the centroid frequency of the LFQPOs do not change significantly with energy, while the full width at half maximum (FWHM) and fractional rms show a complex evolution with energy. 
The LFQPO phase lags at high energies and low energies show consistent energy-dependence relations taking the $\sim$ 2\,keV as reference. Our results suggest that the LFQPOs from high energy come from the LT precession of the relativistic jet, while the low-energy radiation is mainly from the perpendicular innermost regions of the accretion disk.

\end{abstract}



\section{Introduction} \label{sec:intro}

Black hole (BH) X-ray binaries (XRBs) are two-body systems in which a stellar mass BH accretes matter from a companion star.
The accretion process is characterized by a plethora of spectral and variability phenomena that are uniquely observed in BH systems and are related to the strong relativistic nature of BHs.

Typical BH binaries (BHBs) spend most of their time in quiescence states, and become detectable during their outbursts. The outbursts of BHBs are known to pass through a number of spectral states; some of these spectral states are believed to be associated with jet ejection (see review by \citealt{Klis2006} and references therein). 
A hysteresis pattern is often observed during outbursts of BHBs
, which involves the system cycling from a low hard states (LHS), through the hard and soft intermediate states (HIMS, SIMS), the soft
states (SS), and eventually returning back to the LHS (see \citealt{Belloni2010} and \citealt{ Motta2016} for reviews). 
The LHS is characterized by an energy spectrum dominated by a power-law (Comptonized) component with a spectral photon index of $\Gamma \sim1.5-1.7$ and a high energy cutoff around $\sim$100\,keV, and sometimes is supplemented by a weak thermal component. 

The hard X-ray emission is thought to arise from the inverse Compton (IC) scattering of disk photons in a hot and static plasma (often referred to as the corona) \citep{Zdziarski2003} or from the IC scattering in the base of the jet. In contrast, the spectrum of the SS is characterized by a black body disk component, with a typical inner disk temperature of $\sim$1\,keV and an inner disk radius extending to the innermost stable circular orbit (ISCO) of the BH \citep{Steiner2010}. The soft state usually refers to a high mass transfer rate and a high Eddington-scaled luminosity.

In the LHS, strong (up to 50\% for total rms) aperiodic variability is commonly seen in their Fourier power density spectra (PDS), known as the band-limited noise, which is flat at low frequencies and steep at higher frequencies. Compared to the LHS, the total fractional rms is small in the HIMS and SIMS, while drops to a few percent in the SS \citep{Mendez1997,Belloni2005}. During the LHS and HIMS, quasi-periodic oscillations (QPOs) are usually observed in their PDS. QPOs are a common characteristic of accreting BHBs, and the centroid frequencies of low-frequency QPOs (LFQPOs) range from a few mHz to $\sim$30\,Hz \citep{Belloni2002, Casella2004, Motta2015}. The LFQPOs in BHBs are classified into three types, named type-A, type-B, and type-C, based on their quality factor Q, fractional rms, noise component and phase lag (\citealt{Casella2005}). 

Despite LFQPOs being discovered for several decades, their origin is still under debate, and their physical nature is disputed. In particular, their associations with various spectral states suggest that they could be a key element in understanding the physical mechanisms that give birth to the different states. Several models have been proposed to explain the origin of LFQPOs in X-ray binaries, and they generally consider either instability in the accretion flow \citep{Tagger1999, Cabanac2010, Molteni1996} or a geometric oscillation of  accretion flow or a small-scaled jet \citep{Ingram2009, MA2021}. The study of energy dependence of the LFQPO properties, such as fractional rms, centroid frequency and the phase-lag \citep{Rodriguez2004, Casella2004, Qu2010, Yadav2016, Motta2015, Ingram2016, VandenEijnden2017, Zhang2017,  Ingram2020, Karpouzas2021}, has provided crucial information in understanding the origin of the variability and its association with the accretion geometry. 
Recently,  \citet{Karpouzas2020} and  \citet{Bellavita2022} proposed a time-dependent down- and up-Comptonization model that explain the energy-dependent rms and time-lag spectra of QPOs in neutron star systems and BH systems.      This model has successfully interpreted the QPOs in BH GRS 1915+105 \citep{Karpouzas2021,Mendez2022},  MAXI J1348-630 \citep{Garcia2021} and MAXI J1535-571 \citep{zhang2022}. However, the origin of QPOs is not specified in the models of \citet{Karpouzas2020} and  \citet{Bellavita2022}.

MAXI J1820+070 is an X-ray transient discovered on 2018 March 11 by \emph{MAXI}/GSC \citep{Kawamuro2018}. Optical follow-up observations identified a possible optical counterpart coinciding with ASASSN-18ey discovered by ASAS-SN project \citep{Denisenko2018}, while \citet{Kennea2018} identified the same counterpart with \emph{Swift}/UVOT. A radio counterpart was reported by \citet{Bright2018} with the Arcminute Microkelvin Imager Large Array (AMI-LA). 
\citet{Torres2019} confirmed the black hole nature of the compact object, and estimated the mass of the BH to be 8.48$^{+0.79}_{-0.72}$M$_{\bigodot}$ \citep{Torres2020}. The source is located at a distance of 2.96$\pm$0.33\,kpc, and the jet inclination angle is $62\pm3^{\circ}$ \citep{Atri2020}. The spin is constrained to be 0.2$^{+0.2}_{-0.3}$ \citep{Guan2021,zhao2021}.   
The outburst of MAXI J1820+070 in 2018 lasted about eight months. In the bright hard state, \citet{Kara2019} proposed a lamppost geometry, suggesting a contracting corona while the inner radius of the accretion disk remains unchanged with NICER observations. \citet{Buisson2019} compared the evolution of the X-ray spectral and variability with NuSTAR observation and found that the inner disk radius stays constant when the corona moves close to the BH. \citet{YOU2021} proposed a jet-like corona scenario, in which the X-ray emission becomes more concentrated and the corona approaches the BH when the coronal material outflows faster in the hard state. A state transition from hard to soft occurred in early July 2018 \citep{Homan2018b,Homan2018c}; \citet{Wang2021} studied the spectral-timing evolution during the hard-to-soft transition and suggested that the corona expands vertically and triggers the launch of a jet knot that propagates along the jet stream. The soft state lasted for $\sim$3 months, and from September 2018 the source began its transition to the hard state \citep{Homan2018d} again.

LFQPOs with a centroid frequency at several tens of mHz have been observed in both the X-ray and optical bands during the initial LHS of the outburst of MAXI J1820+070 \citep{Buisson2018,Homan2018a,Mereminskiy2018,Yu2018a,Yu2018b,Zampieri2018}. It is worth noticing that LFQPOs have been detected above 200 keV and up to 250 keV by \emph{Insight}-HXMT in the hard state \citep{MA2021}. 
\citet{Stiele2020} studied the four outbursts from March 2018 to October 2019 using  Swift/XRT and NICER data. Type-C QPOs with frequencies below 1\,Hz have been observed in most of the observations, suggesting that the source was in a low effective accretion state for most part of the outburst. During the transition from the hard to soft state in early July 2018, type-C QPOs show a rapid increase in frequency and a switch from type-C to type-B QPOs. In addition, a strong radio flare was shortly observed $\sim$1.5--2.5\,hr after the switch, suggesting a strong association with the discrete jet ejection \citep{Homan2020}.


Studying the evolution of rapid X-ray variability throughout the outbursts of BHs helps us better understand the evolution of the accretion disk/corona geometry. Meanwhile, the energy-dependent properties of LFQPOs could provide valuable information on the physical mechanisms of their origins. In this paper, we focus on the study of the energy-dependent QPO properties from the 2018 outburst of MAXI J1820+070 using \emph{Insight}-HXMT and NICER observations, extending our studies of LFQPOs to a broad energy range of 0.2 -- 200\,keV, to comprehensive understand the physical nature of QPOs. We present the observations and data reduction in Section 2, the main results in Section 3, and the discussion and conclusion in Section 4 and Section 5.

\section{Observations and Data Analysis} \label{sec:obs}

\begin{figure}
\begin{minipage}{0.45\textwidth}
\includegraphics[height=0.8\linewidth,angle=0]{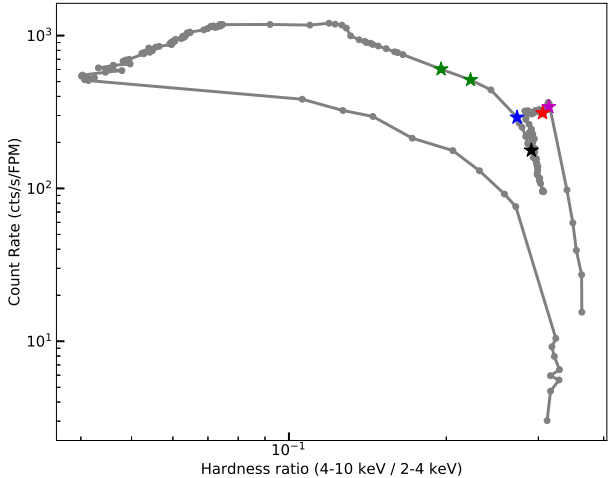}
\end{minipage}
\caption{NICER hardness-intensity diagram (HID) of MAXI J1820+070. Intensity is the counts rate in 0.5-10.0\,keV. The hardness ratio is the ratio of the counts rates from 4.0-10.0\,keV and 2.0-4.0\,keV bands. The observations selected for timing analysis are highlighted by stars.
\label{fig:hid}}
\end{figure}
\subsection{NICER}

NICER observed MAXI J1820+070 for a total of 750 individual visits between March and November 2018. 
We generate lightcurves using \emph{xselect}, without subtracting the background since the background contribution can be ignored (typically less than 2 counts s$^{-1}$)
because of the very high count rate of this source.

In Figure \ref{fig:hid}, we show the NICER hardness intensity diagram (HID) of the full outburst. The hardness ratio was defined as the ratio of the count rates between the 4-10 keV and 2-4 keV energy bands. In order to study the property of the broad energy band QPOs, we try to choose the simultaneous NICER and  \emph{Insight}-HXMT observations. 
The observations selected for timing analysis are highlighted by stars, and the log is listed in Table \ref{table:obs}. During the HIMS (ObsID: 1200120196), the QPO frequency evolves very quickly, from 2\,Hz to 4.2\,Hz, and we thereby split the observation into several time intervals.

\begin{table}
\footnotesize
\caption{Log of NICER and \emph{Insight}-HXMT Observations of MAXI J1820+070 during its 2018 outburst.}
\label{table:obs}
\medskip
\begin{center}
\begin{tabular}{lllll}
\hline \hline
Date & Telescope & MJD & ObsID & Frequency   \\
&&&&(Hz) \\
\hline
2018-03-22 & NICER & 58199.026 & 1200120107  & 0.038$\pm$0.001 \\
&HXMT& 58199.450 & P0114661003  & 0.038$\pm$0.001 \\
\hline
2018-04-16 & NICER & 58224.068 & 1200210130 & 0.120$\pm$0.002 \\
\hline
2018-06-28 & NICER & 58297.194 & 1200120189 & 0.419$\pm$0.011 \\
& HXMT & 58297.244 & P0114661078 & 0.423$\pm$0.002 \\
\hline
2018-07-03 & NICER & 58302.068 & 1200120194 & 0.775$\pm$0.012\\
\hline
2018-07-05 &NICER & 58304.272 & 1200120196(a)  & 2.073$\pm$0.023 \\
\hline
2018-07-05 &NICER & 58304.640 & 1200120196(b)  & 3.010$\pm$0.018 \\

\hline
\hline
\end{tabular}
\end{center}
\end{table}

\subsection{\emph{\emph{Insight}}-\emph{HXMT}}

\emph{Insight}-HXMT is the first X-ray astronomy satellite of China, and was successfully launched on June 15, 2017 \citep{zhang2020}.
\emph{Insight}-HXMT carries three slat-collimated main payloads
onboard: the High Energy X-ray telescope (HE, NaI(Tl)/CsI(Na), 20-250\,keV, 5100\,cm$^{2}$), the Medium Energy X-ray telescope (ME, Si-PIN, 5-30\,keV, 952\,cm$^{2}$), and the Low Energy X-ray telescope (LE, SCD, 1-15\,keV, 384\,cm$^{2}$). Each payload includes blind FoV (Field of View) detectors used to estimate the particle induced instrumental background.

\emph{Insight}-HXMT monitored MAXI J1820+070 from March 2018 to October 2018 with over 140 pointing observations. Since the hard to soft state transition happened in a very short time, \emph{Insight}-HXMT missed the hard-intermediate state of the outburst. Therefore, we study the state transition only with NICER observations. 
We use the \emph{Insight}-HXMT Data Analysis software (HXMTDAS) v2.05 
to analyze all the data. We filter the data with the following criteria:
(1) pointing offset angle $< 0.05^{\circ}$; (2) elevation angle $> 6^{\circ}$;
(3) the value of the geomagnetic cutoff rigidity $> 6$\,GeV. Only events from the small FoV are selected for analysis. 
The backgrounds are estimated with the background models in \citet{Liao2020a}, \citet{Guo2020} and \citet{Liao2020b}. 


\begin{table}
\footnotesize
\caption{Different energy bands table.}
\label{table:1}
\medskip
\begin{center}
\begin{tabular}{lll}
\hline \hline
Telescope & Energy band & Centroid Energy  \\
          & (keV)       & (keV)  \\
\hline
        & 0.2--0.4   & 0.3 \\
        & 0.4--0.6   & 0.5 \\
        & 0.6--0.85   & 0.72 \\
        & 0.85--1.2   & 1.02 \\
 NICER  & 1.2--1.75   & 1.48 \\
        & 1.75--2.5   & 2.12 \\
        & 2.5--3.6   & 3.05 \\
        & 3.6--5.2   & 4.4 \\
        & 5.2--7.4   &  6.3\\
        & 7.4--9.4    &  8.4 \\
\hline
        & 1--2.6   & 1.8 \\
LE/HXMT & 2.6--4.8 &  3.7 \\
   & 4.8--7   &  5.9 \\
   & 7--11    &  9 \\
\hline
   & 7--11 &   9\\
ME/HXMT & 11-23 &  15.5 \\
   & 23--35 &   27.5\\
\hline
   & 25--33   &  29 \\
   & 33--48   &   40.5\\
HE/HXMT & 48-67   &  57.3\\
   & 67--100  &   83.4\\
   & 100--150 &   123.3\\
   & 150--200 &   173.3\\
\hline
\hline
\end{tabular}
\end{center}
\end{table}
\subsection{Data Analysis}
Power density spectra (PDS) are calculated using $powspec$ with Miyamoto normalization \citep{Miyamoto1991}, and the Poisson noise is subtracted.
The PDS are geometrically re-binned in increments of 3\% of the frequency.
We use a combination of Lorentzians \citep{Belloni2002} functions to fit the power spectra: two or three Lorentzians for the underlying continuum; one Lorentzian for the fundamental QPO, one for the harmonic, and one for a possible sub-harmonic with the centroid frequency linked to the half of the fundamental QPO frequency. The PDS can be well fitted with no obvious structure in the residuals. 
The errors are given at 90\% confidence region computed from the XSPEC MCMC routine.

The background contribution to the fractional rms is corrected according to rms = $\sqrt{P}*(S+B)/S$, where $S$ and $B$ stand for source and background count rates, respectively, and $P$ is the power normalized according to \citet{Miyamoto1991}.

In order to study the energy-dependent LFQPO properties, we divide the whole energy band of HXMT and NICER into 23 sub-channels to create the corresponding light curves and power spectra. The selection of energy channels and the corresponding energy ranges are listed in Table \ref{table:1}.

Frequency-dependent phase-lag spectra between a soft band and a hard band are also created, following the method described in \citet{Vaughan1997} and \citet{Nowak1999}, via $stingray$ \citep{Huppenkothen2019}. To quantify the phase-lag behaviour of the QPOs, we extract the phase lags in a frequency range centered at the QPO peak frequency and spread to the width of the peak ($\nu$ $\pm$ FWHM/2). Positive lags mean hard photons lag the soft ones.

\begin{figure*}[htbp]
\centering
\begin{minipage}[t]{0.48\textwidth}
\centering
\includegraphics[height=0.7\linewidth]{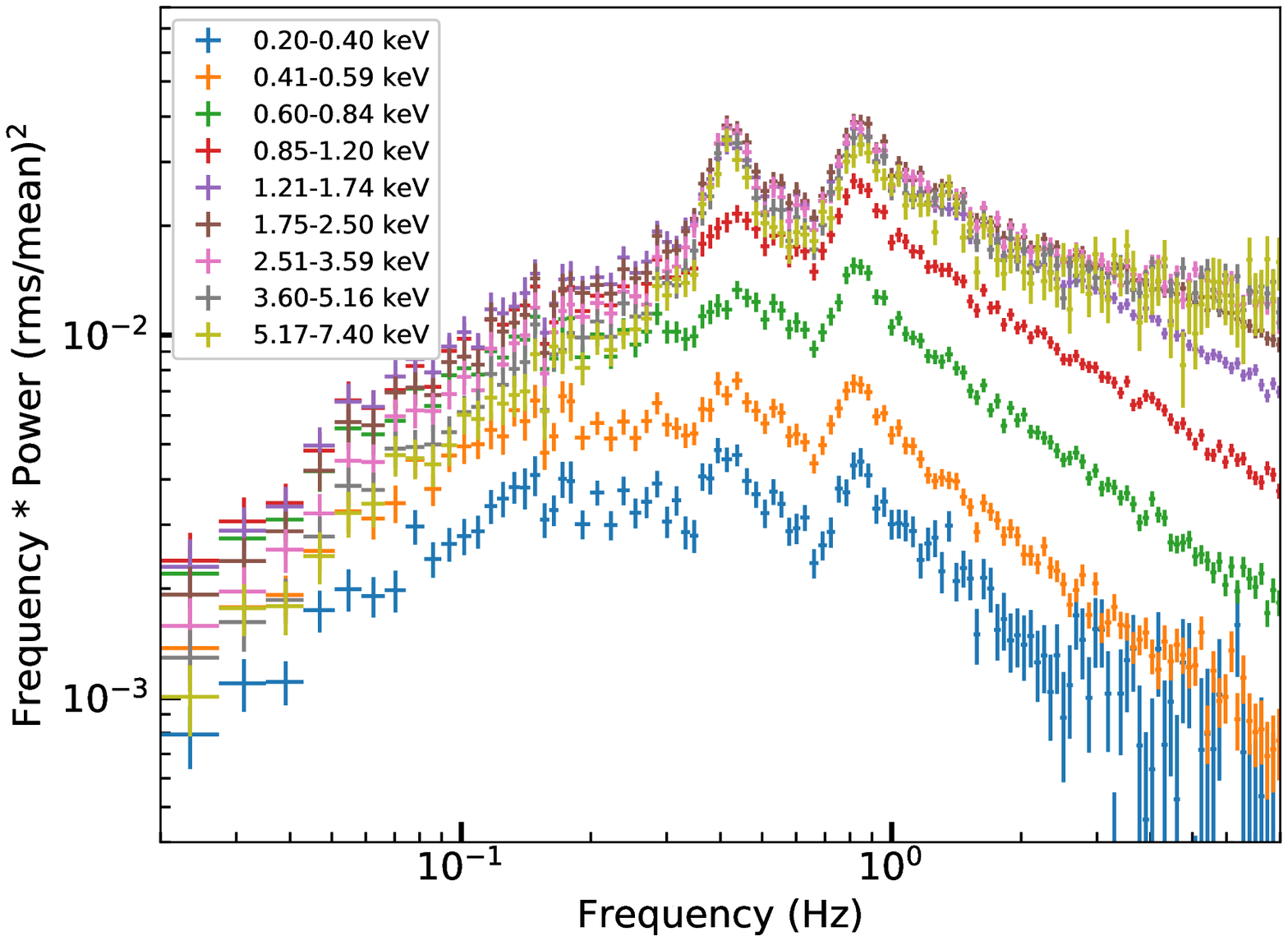}
\end{minipage}
\begin{minipage}[t]{0.48\textwidth}
\centering
\includegraphics[height=0.7\linewidth]{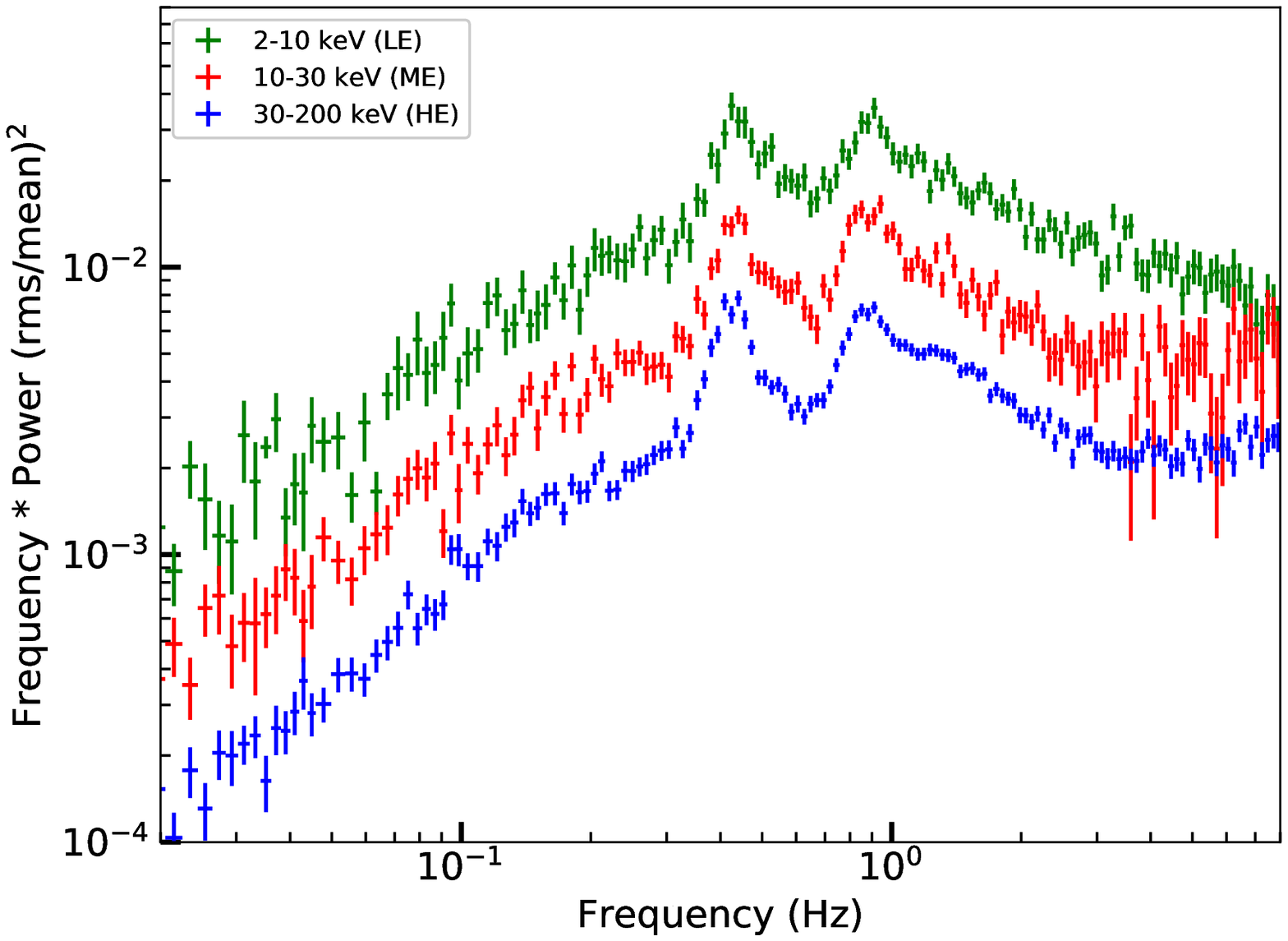}
\end{minipage}
\caption{Left: Representative PDS at different energies within 0.2-7.4\,keV (NICER, obsID:1200120189). Right: Representative PDS of 2-10\,keV (\emph{Insight}-HXMT/ME), 10-30\,keV (\emph{Insight}-HXMT/ME), and 30-200\,keV (\emph{Insight}-HXMT/HE) (obsID: P0114661078). The QPOs are detected at $\sim$0.4\,Hz in all sub-energy bands.
\label{fig:f4}}
\end{figure*}

\section{Results} \label{sec:results}
\subsection{Power Density Spectra} \label{subsec:pds}
At the early stage of the outburst, when the source flux increases quickly, the PDS are dominated by three band-limited noise components, which are commonly observed in black holes during the LHS (e.g. GX 339-4, \citealt{Belloni2010}). Later, the type-C LFQPOs began to be continuously observed by \emph{Insight}-HXMT, from MJD 58194 to MJD 58301.
Since the source does not show significant high-frequency phenomena, we focus our study on the behavior of LFQPOs. 

Figure \ref{fig:f4} shows representative PDS in different energy bands from the nearly simultaneous observations of NICER and \emph{Insight}-HXMT on MJD 58297. The left panel show PDS from all NICER sub-energy channels in 0.2--7.4\,keV energy bands (see table \ref{table:1} for detail, the PDS in 7.4-9.4\,keV is not included because of the very large error bars), and the right panel gives the PDS from \emph{Insight}-HXMT/LE (2-10\,keV), \emph{Insight}-HXMT/ME (10-30\,keV) and \emph{Insight}-HXMT/HE (30-200\,keV), separately. In all PDS, a strong type-C QPO at a frequency of $\sim$0.42\,Hz is observed, accompanied by a flap-top noise component dominating below 0.1\,Hz. 
The FWHMs are relatively higher at low energy bands ($<$1.2\,keV). A second-harmonic peak at around 0.85\,Hz is also detected.
The PDS of NICER show very similar shapes, while the shape of \emph{Insight}-HXMT PDS at high frequency evolves with energy. The detailed properties of the broadband noise have been studied in \citet{Yang2022}, and are thus excluded from our work.


\subsection{Energy dependence of QPO parameters } \label{subsec:d2}

\begin{figure}
\plotone{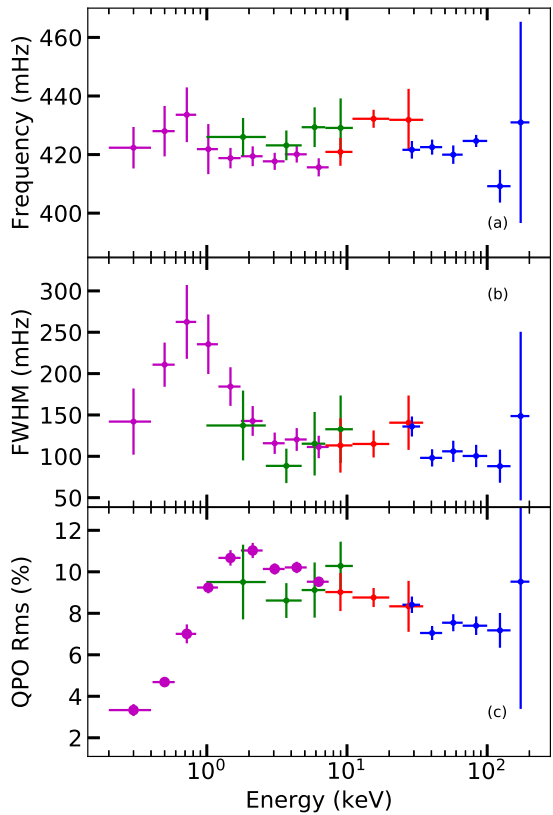}
\caption{Energy dependence of the QPO centroid frequency (a), FWHM (b), and fractional rms (c) for observations of MAXI J1820+070 using combined NICER and \emph{Insight}-HXMT data. The purple, green, red and blue points represent NICER, \emph{Insight}-HXMT/LE, \emph{Insight}-HXMT/ME and \emph{Insight}-HXMT/HE data, respectively.
\label{fig:qporms}}
\end{figure}

In order to quantitatively study the energy dependence of the QPO properties, we study power spectra from different energy bands listed in table \ref{table:1}. 
Type-C QPOs are detected by \emph{Insight}-HXMT from MJD 58194 to MJD 58301, while the QPO parameters show very similar behaviors with energy. In Figure \ref{fig:qporms}, we show the energy-dependent properties of a representative QPO observed in the energy band of 0.2\,keV to 200\,keV by NICER and \emph{Insight}-HXMT on MJD 58297 in the hard state.

The centroid frequency of the QPO does not change significantly with energy, as shown in Figure \ref{fig:qporms}(a). However, the FWHM of the QPO shows a complex evolution with energy below 2\,keV: the FWHM first increases with energy till $\sim$1\,keV, then gradually decreases with energy till $\sim$2\,keV, and becomes nearly constant above $\sim$2\,keV. The QPO fractional rms monotonically increases from $\sim$3\% to 11\% with energy in 0.2\,keV $\sim$2\,keV band, beyond which the QPO rms slightly decreases with energy. 


\subsection{Phase Lag} \label{subsec:phaselag}
\begin{figure}
\plotone{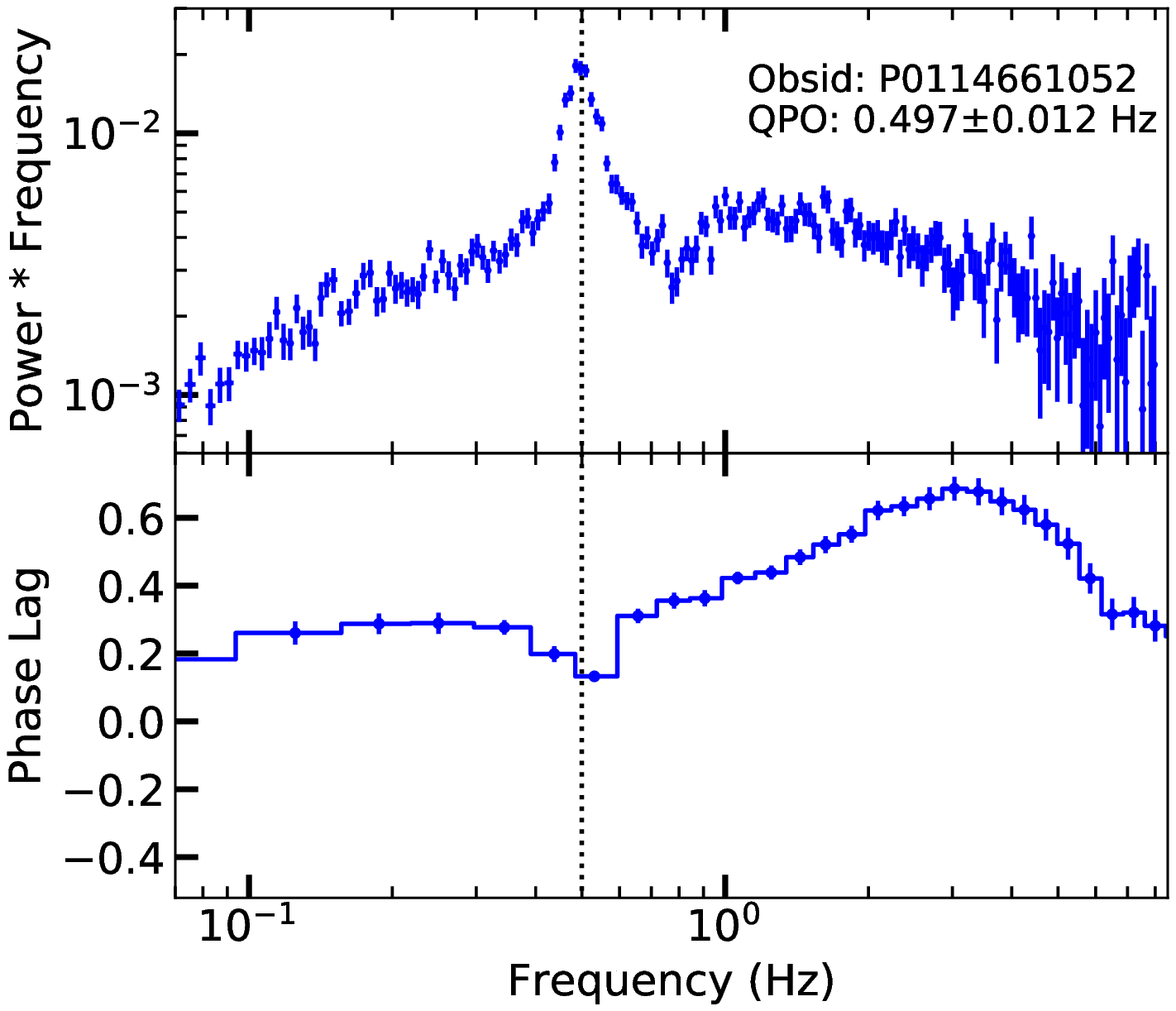}
\plotone{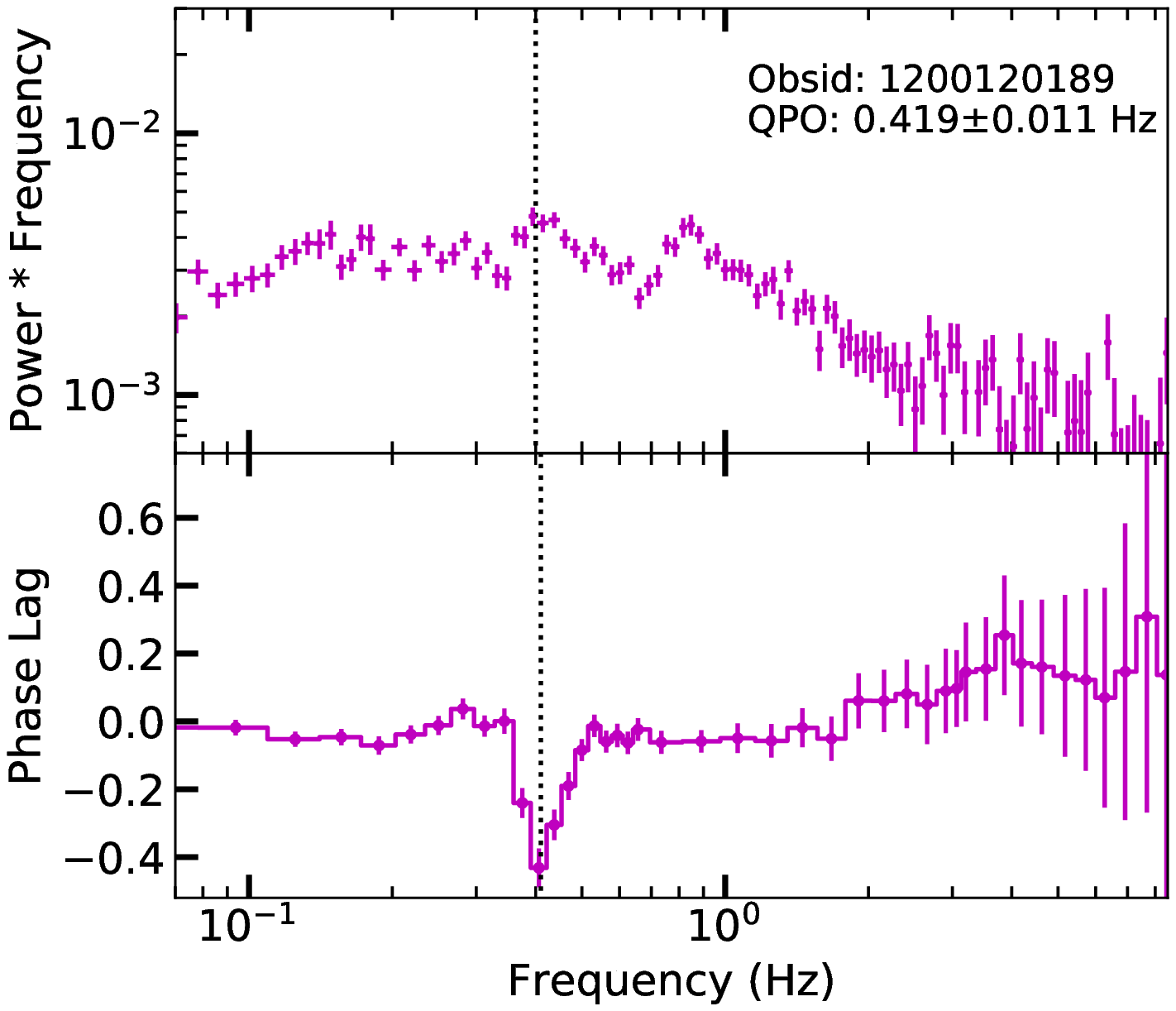}
\caption{Examples of phase-lag spectra as a function of Fourier frequency, from \emph{Insight}-HXMT/HE and NICER. Upper: The phase lags spectra are calculated for the \emph{Insight}-HXMT/HE 67--100\,keV bands relative to the 1--2.6\,keV. Bottom: The phase lags spectra are calculated for the 0.2--0.4 \,keV bands relative to the 1.2--1.75\,keV. The vertical dotted lines indicate the QPO fundamental frequency.
\label{fig:phaselag_HE}}
\end{figure}

\begin{figure*}
\centering
\begin{minipage}{0.32\textwidth}
\includegraphics[height=0.6\linewidth]{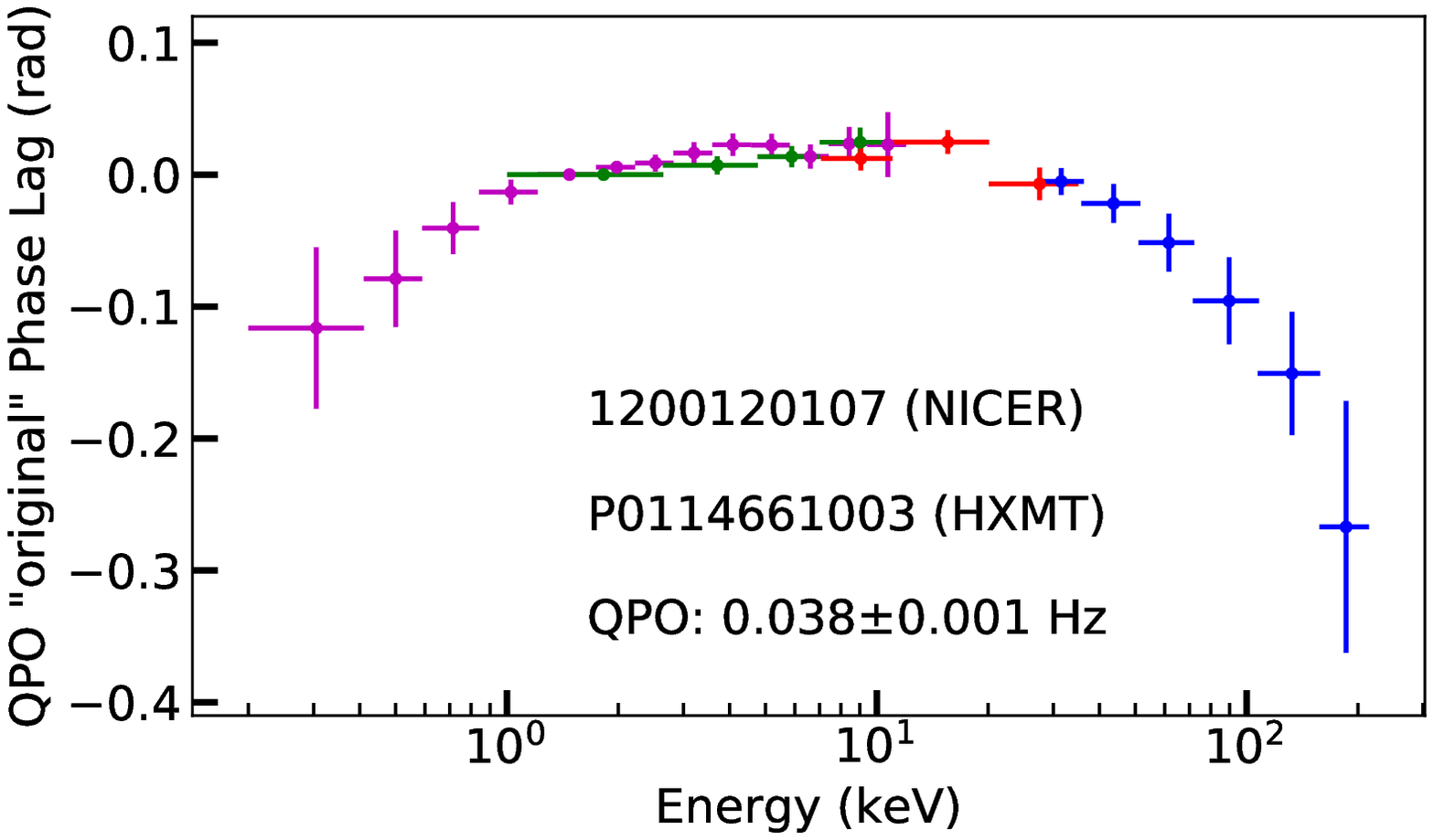}
\end{minipage}
\begin{minipage}{0.32\textwidth}
\includegraphics[height=0.6\linewidth]{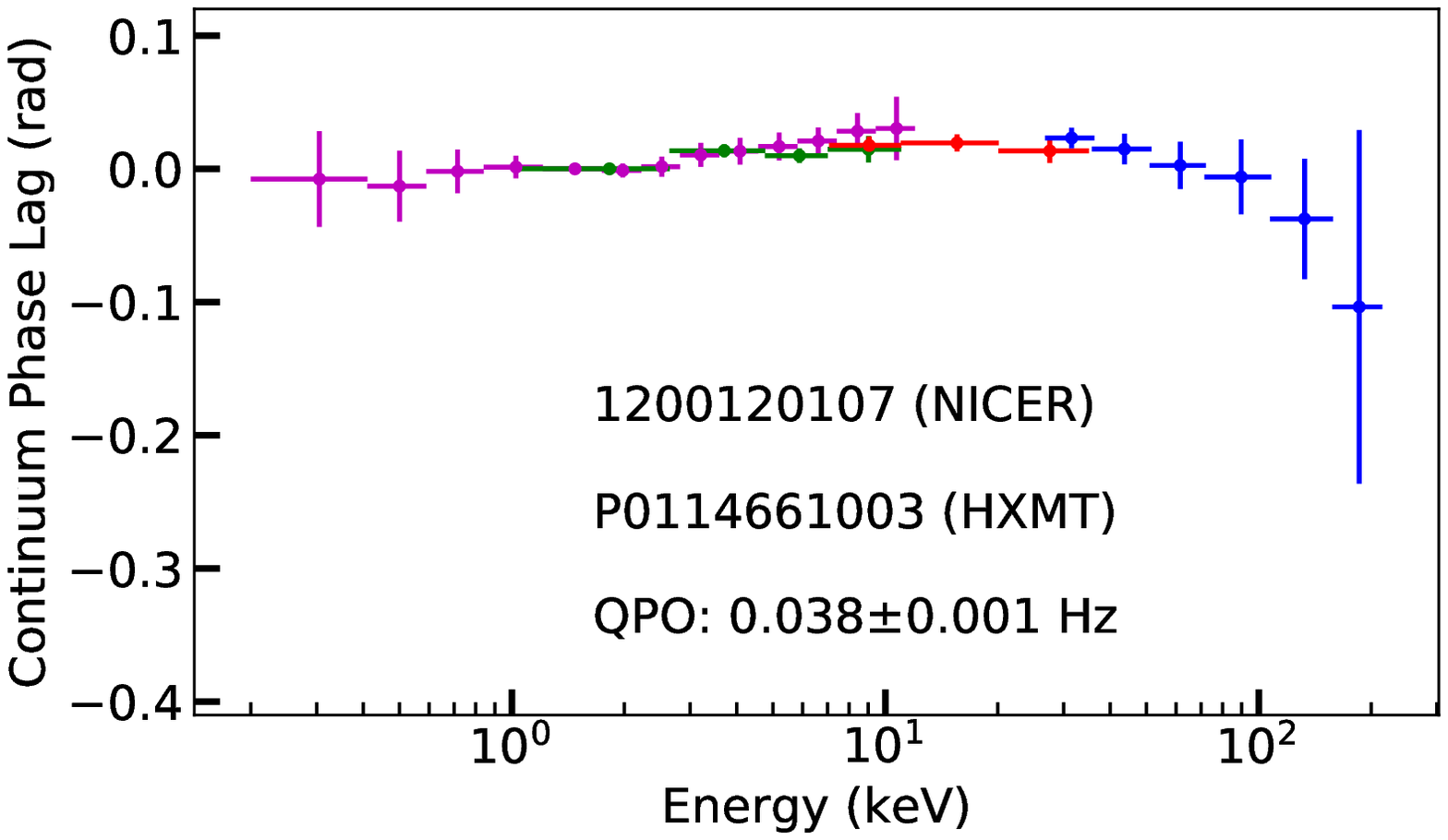}
\end{minipage}
\begin{minipage}{0.32\textwidth}
\includegraphics[height=0.6\linewidth]{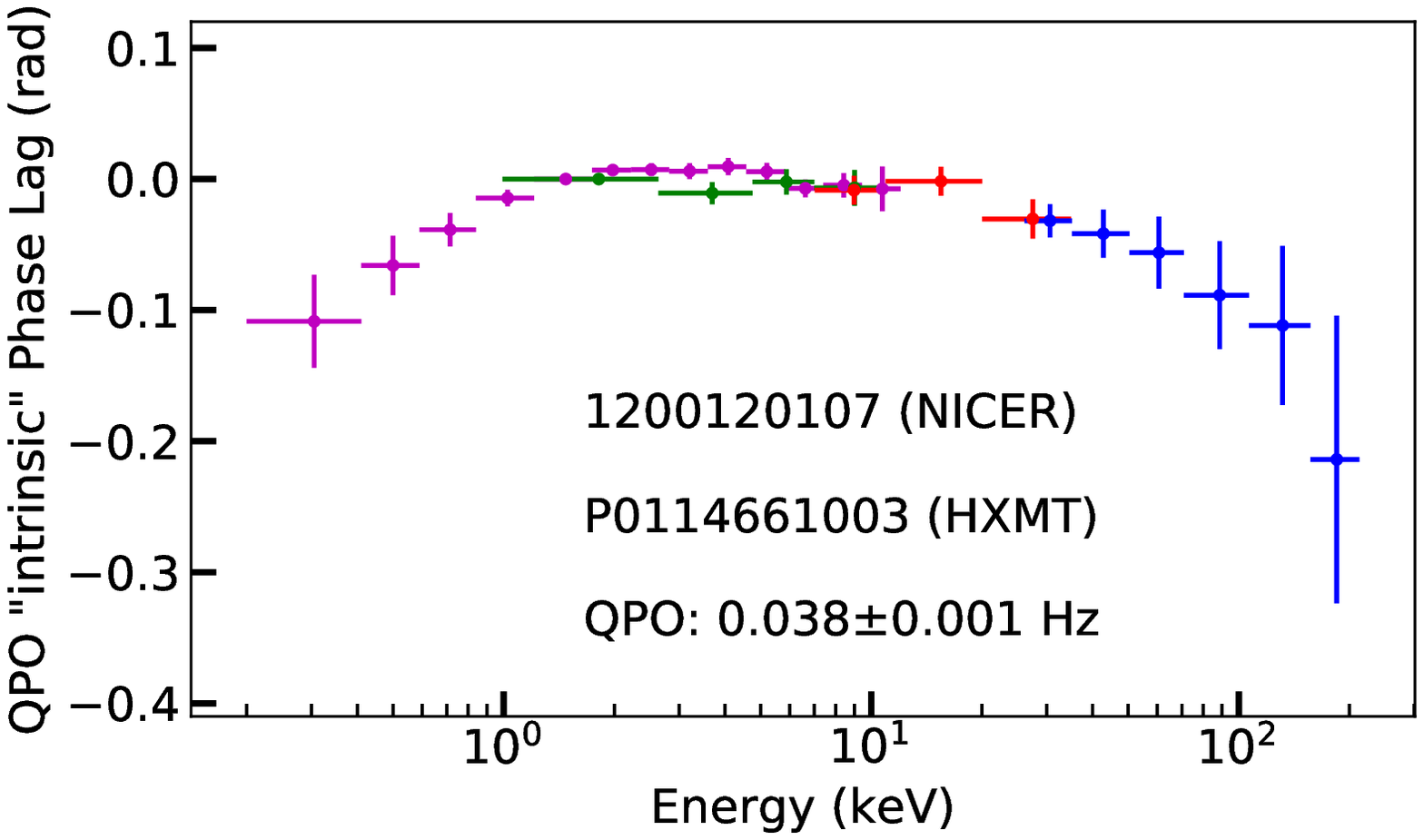}
\end{minipage} \\
\begin{minipage}{0.32\textwidth}
\includegraphics[height=0.6\linewidth]{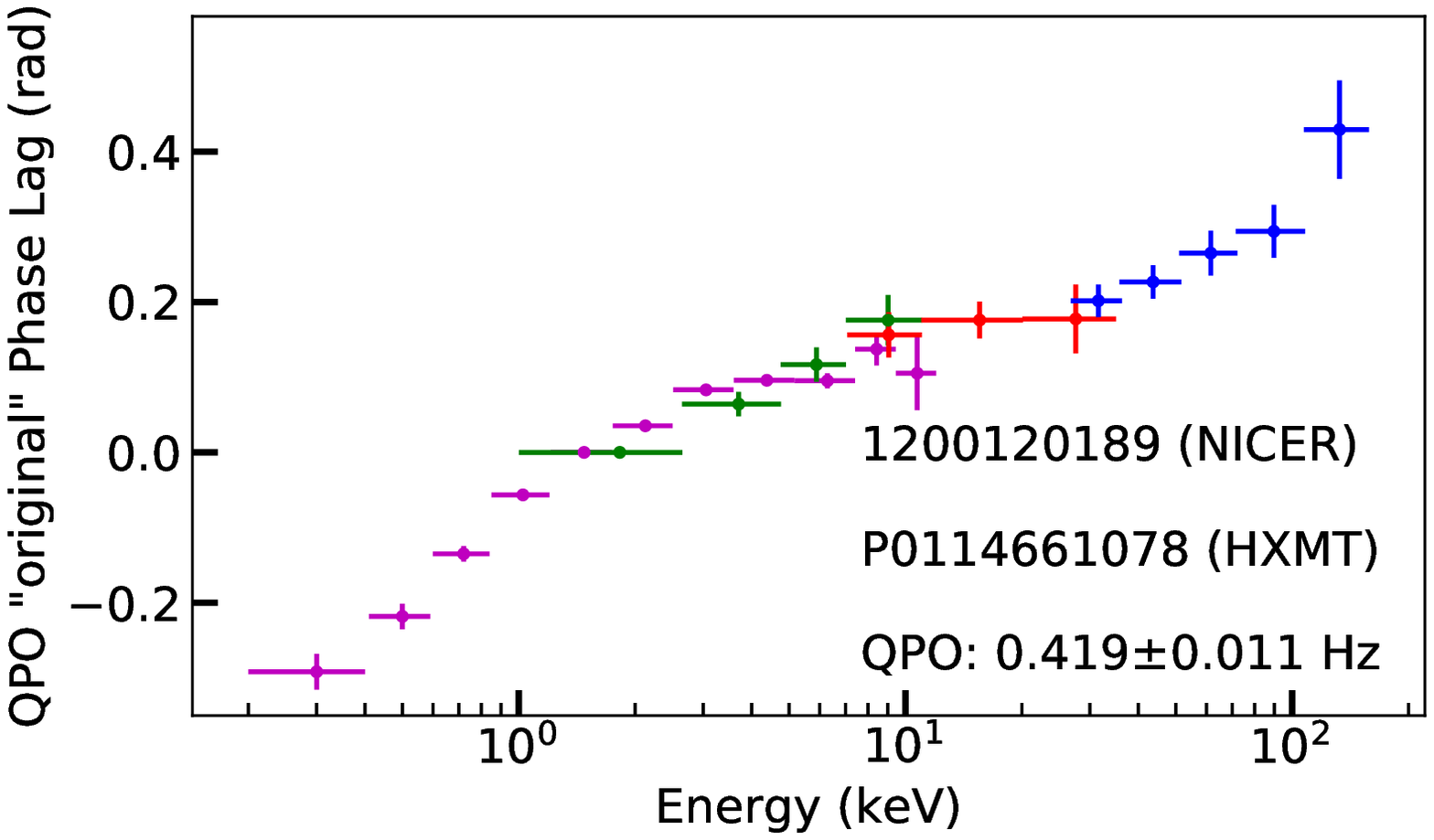}
\end{minipage} 
\begin{minipage}{0.32\textwidth}
\includegraphics[height=0.6\linewidth]{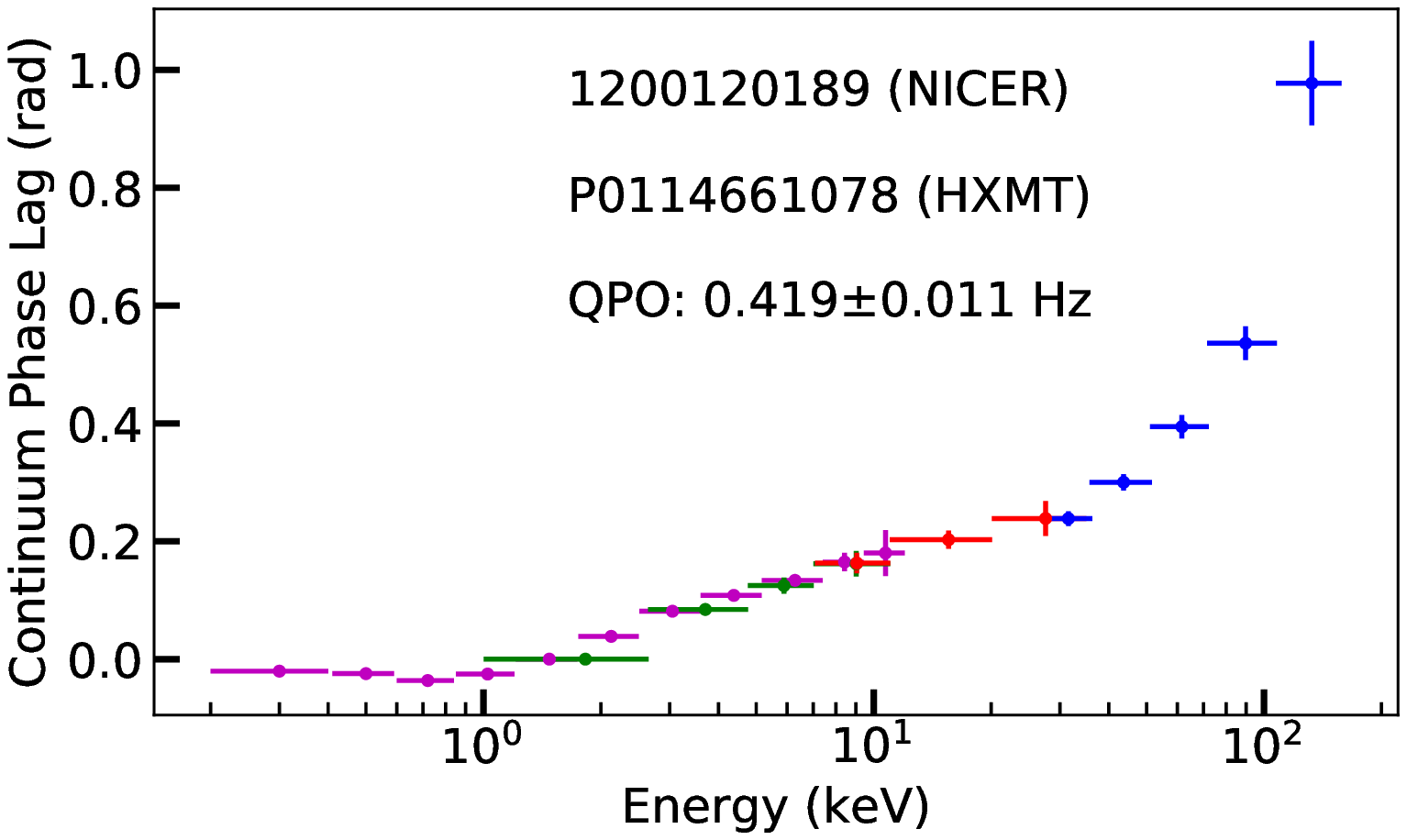}
\end{minipage} 
\begin{minipage}{0.32\textwidth}
\includegraphics[height=0.6\linewidth]{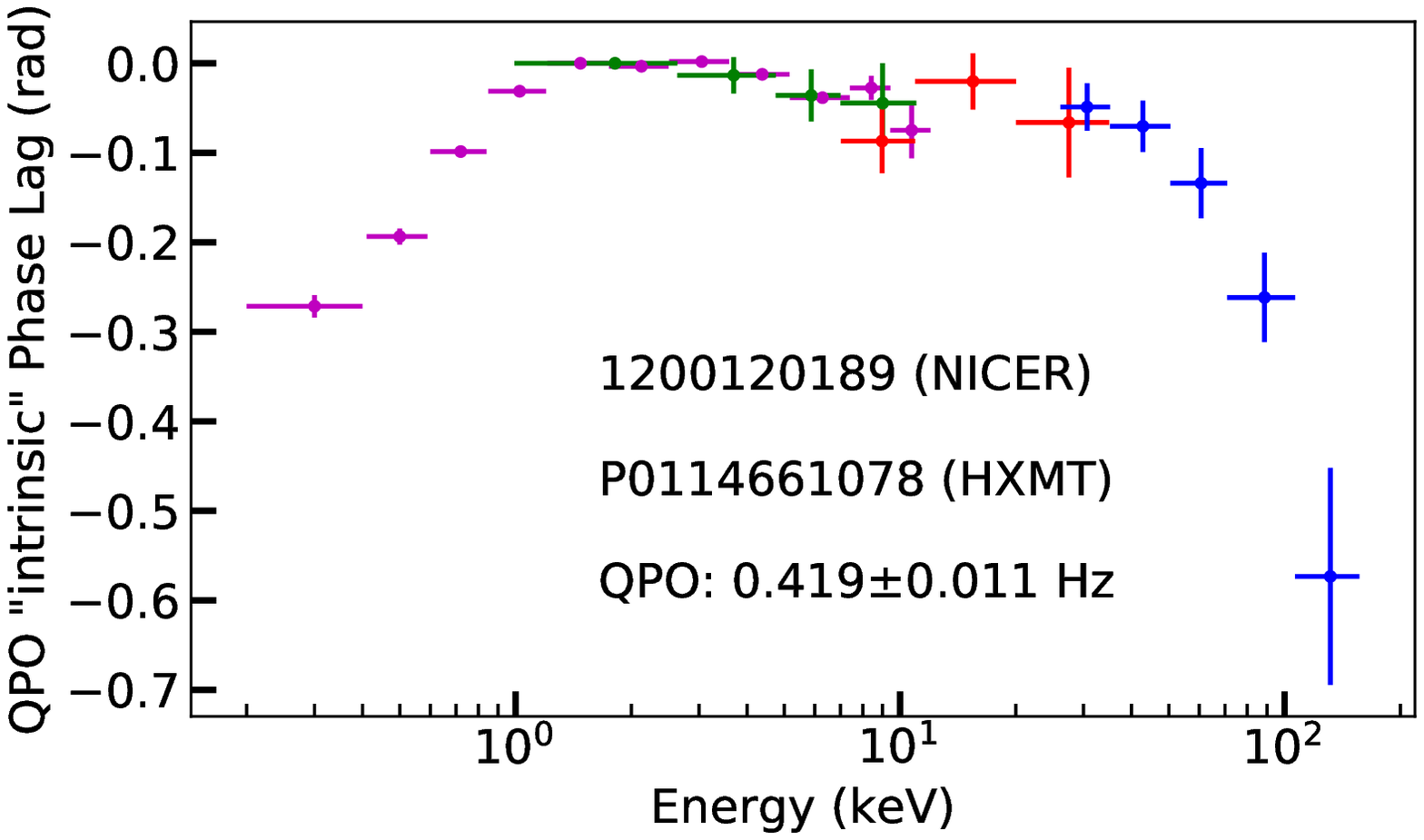}
\end{minipage}

\caption{QPO phase lags as a function of photon energy. From top to bottom, the ``original'' QPO phase lags , the phase lag of the continuum and the ``intrinsic'' QPO phase lags. The purple, green, red and blue points represent NICER, \emph{Insight}-HXMT/LE, \emph{Insight}-HXMT/ME and \emph{Insight}-HXMT/HE data, respectively.
\label{fig:qpo-phaselag}}
\end{figure*}

Figure \ref{fig:phaselag_HE} shows representative frequency-dependent phase-lag spectra from \emph{Insight}-HXMT and NICER. 
A dip-like feature centered at the QPO centroid frequency is always significantly present in the lag-frequency spectra. Above the QPO frequency, the phase lag first increases rapidly to its maximum, and then gradually decreases with Fourier frequency. This dip-like feature seems to be more obvious when the QPO frequency increases \citep{MA2021}, and is not always centered on, but sometimes deviates slightly from the QPO centroid frequency. Such features have been reported in other BHBs, such as GRS 1915+105 \citep{Yadav2016} and GX 339-4 \citep{Zhang2017}, but with a much shorter time scale. 
We further find that the lag-frequency spectra computed from different sub-energy bands are very similar, while the amplitude of the phase lag at the QPO frequency increases with energy.


The dip at the LFQPO frequency in the lag-frequency spectra is a common characteristic in MAXI J1820+070, suggesting that the feature is intrinsically associated with the LFQPO. Thus the production of the dip may be exclusively connected to the physical mechanism producing LFQPO, so we consider the depth of the dip as the ``intrinsic'' phase lag for QPOs. Other high energy processes, such as inward propagation of fluctuation in accretion flow, may contribute to the broadband noise in PDS and the phase-lag continuum, which may enhance the ``continuum" of the lag-frequency spectra. We, therefore, need to remove the phase-lag continuum from the lag directly measured at QPO frequency which is referred to as ``original'' phase lag. Considering the lag-frequency spectrum is flat below the QPO frequency, we use the average phase-lag value below the QPO frequency as the continuum phase-lag. \citet{Zhou2022} proposed a convolution mechanism between the strong broadband noise and QPOs in the time domain, and their phase lags could be obtained by linear correlation. Based on their works, 
we calculate the QPO ``intrinsic'' phase lag by subtracting the phase-lag continuum from the ``original'' phase lag.
The uncertainty is estimated by propagating the errors of all parts.  

Figure \ref{fig:qpo-phaselag} shows the ``original'' QPO phase lag, continuum phase and ``intrinsic'' phase lag at different energy bands for combined NICER and \emph{Insight}-HXMT observations.
The ``original'' QPO phase lags show similar energy dependencies below $\sim$10\,keV, in which the soft lags increase with energy, while show different energy dependencies above 10\,keV, in which the soft lags either decrease or increase with energy. While the ``intrinsic'' QPO phase lags show similar energy dependencies. In particular, the phase lags are negative below 2\,keV and above 10\,keV.

\subsection{QPO Properties in different states with NICER Observations}

\begin{figure}[htbp]
\includegraphics[width=0.47\textwidth]{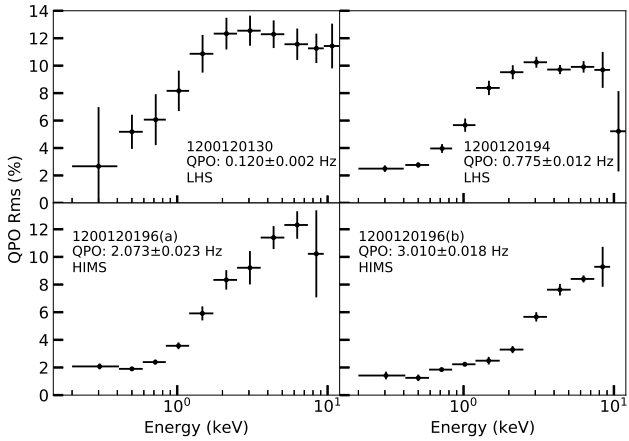} 
\caption{The QPO fractional rms computed from different energy bands using NICER observation.
\label{fig:qporms_nicer}}
\end{figure}

\begin{figure}
\includegraphics[width=0.49\textwidth]{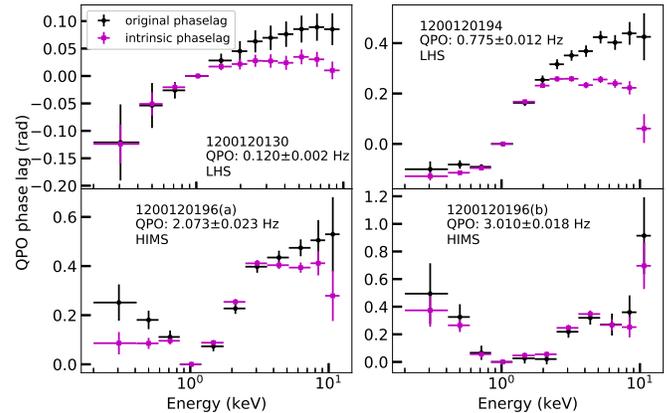}
\caption{The QPO phase lag at different energy bands using NICER Data. The black and purple points represent the ``original'' and ``intrinsic'' QPO phase lag.
\label{fig:qpolag_nicer}}
\end{figure}

Considering the lack of \emph{Insight}-HXMT observations during the HIMS, we use NICER data only to study the QPO properties during the state transition, particularly on the comparison of the energy dependence of QPO rms and phase lag between the hard and the hard-intermediate states.
In the LHS (ObsID: 1200120130 and 1200120194 in Figure \ref{fig:qporms_nicer} and \ref{fig:qpolag_nicer}), the QPO rms increases from $\sim$3\% at 0.2\,keV to about 13\% around 2\,keV, and then stays more or less constant above 2\,keV. While the QPO phase lag increases monotonically with phonon energy from 0.2\,keV to $\sim$2\,keV, then becomes constant near 0.
During the HIMS (ObsID: 1200120196), the QPO rms stays almost constant $<$1\,keV and then increases monotonically with energy up to $\sim$10\,keV, though the highest energy band has very large error bars. The rms-energy relations are distinctly different from those in the LHS, while the break energy (where the rms-energy relation dependence change from increasing to flat) increases with QPO frequency. The QPO phase lags-energy relations show turn-over trends, which show positive lags below 0.8\,keV and above 1.2\,keV when the QPO frequency is higher than 1\,Hz. 

\section{DISCUSSIONS and summary} \label{sec:discussion}

We have reported the timing analysis of the black hole candidate MAXI J1820+070 during its 2018 outburst using \emph{Insight}-HXMT and NICER observations. Type-C QPOs are observed during the hard and the hard intermediate states of the outburst, from MJD 58194 to MJD 58304. For the first time we studied the fractional rms, the FWHM, the centroid frequency, and the phase lag of the type-C QPOs as a function of the photon energy in the 0.2 -- 200\,keV band.
In the LHS, the QPO rms amplitude first rapidly increases with energy from 0.2 to $\sim$2\,keV, then slightly decreases or stays more or less constant above $\sim$2\,keV (Figure \ref{fig:qporms}). The QPOs show soft lags below and above 2-10\,keV (Figure \ref{fig:qpo-phaselag}). While in the HIMS, the QPO rms amplitude increases with energy from 0.2 to 7\,keV or above (Figure \ref{fig:qporms_nicer}), and the QPO show hard lags relative to 0.8-1.2\,keV below 10\,keV (Figure \ref{fig:qpolag_nicer}). The QPO frequency almost keeps constant throughout the outburst, and the FWHM shows a complicated energy dependence. 

The QPO rms-energy relations found in the HIMS is similar to what had been observed in XTE J1550-564, XTE J1650-500 \citep{Rodriguez2004, Gierlinski2005}, GX 339-4 \citep{Zhang2017}, and MAXI J1348-630 \citep{Belloni2020}. In these BHBs, the break energy is usually around 10\,keV, while the rms value is near 10\%. In the LHS, the break energy at $\sim$2\,keV has rarely been observed in other black holes. One possible explanation to this change could be caused by an increase of contribution from the dominated variability component at lower energy \citep{Huang2018, Liu2021}.
\citet{You2018} simulated the evolution of the variability during the state transition, where the thermal disk emission dominates the spectrum and the point of the energy dependence of QPO fractional rms changing from increasing to flat moves to higher energies.  \citet{Wang2021} fitted the spectra of MAXI J1820+070 from its hard state and the bright hard-to-soft transition state and found that the disk temperature increases from $\sim$0.3\,keV (hard state) to $\sim$0.6\,keV (hard-to-soft state), which is consistent with the ratio of component of thermal disk becoming significant at higher energy. 
\citet{You2018} simulated the fractional rms of the type-C QPO in the framework of the Lense-Thirring precession model to study the energy-dependent variability during the state transition. The variability from the disk emission, the Comptonized emission and the reflection component was discussed. Their simulation suggested that the disk emission dominates over the others at low energy bands ( $E<$1\,keV), while the variability is mainly affected by the changing appearance of the occulted area of the disk and the intrinsic blueshift/redshift modulation of the rotating disk. However, the simulation assumes a truncated disk geometry instead of a lamppost geometry. 

The QPO frequency changing with energy had been reported in GRS 1915+105 \citep{Qu2010}, XTE J1550-564 \citep{Li2013a}, H1743-322 \citep{Li2013b}, and XTE J1859+226 \citep{VandenEijnden2017}. 
It is notable that in most cases, the frequency drift becomes significant when the QPO frequency is high \citep{VandenEijnden2017}.  
For MAXI J1820+070, the centroid frequency of type-C QPOs does not change significantly with photon energy, implying that the same QPO frequency is given at different radiation regions that could  include different parts (i.e. different photon energies) of the jet or inner hot flow.




The physical origin of the QPO phase lags is still under debate. The energy dependence of type-C QPO phase lag has been studied in many BHs, such as GRS 1915+105 \citep{Qu2010}, MAXI J1348--630 \citep{Alabarta2022}, GX 339--4 \citep{Zhang2017} and MAXI J1535--571 \citep{2022MNRAS.514.3285G}. We found that the relation between phase lag and photon energy is different with various QPO centroid frequencies (Fig.\ref{fig:qpolag_nicer}). When the QPO frequency is lower than 1\,Hz in the LHS, the QPO phase lag monotonically increases with energy, consistent with the type-C QPOs found in other BHs. However, in the HIMS for the QPO frequency above 1\,Hz, the energy dependence of QPO phase lag shows a turn-over correlation with energy instead of a monotonic evolutionary trend with energy, which is similar to type-B QPO in MAXI J1348-630.   
\citet{Belloni2020} studied the energy dependence of the phase lag at the type-B QPO frequency in MAXI J1348-630, where the photons showed a hard lag between 2--3\,keV and photons at all energy bands using NICER observations. Their result is consistent with the type-C QPO phase lag of MAXI J1820+070 in the HIMS. \citet{Belloni2020} explained the energy dependence of phase lags using Comptonization and showed that a nonphysical large corona is needed to produce observed lags, suggesting that Comptonization coming from a jet is a more realistic model. The energy dependence of QPO phase lags in the LHS is distinctly different from the HIMS, indicating that the origin of the QPO phase lags in the hard state may be different. 

The Lense-Thirring precession of the hot inner flow model is widely used to explain the origin of type-C QPOs in BHs, such as GX 339--4 \citep{Zhang2017}, H1743--322 \citep{Ingram2016} and MAXI J1535--571 \citep{Huang2018}. In this model, a truncated disc geometry is assumed. When the accretion rate increases, the inner disc radius of the disc will be reduced, thus the precession frequency will increase. For MAXI J1820+070, the disk/corona geometry is somewhat mysterious. Based on the modelling of the reflection component and the iron line profile, \citet{Kara2019} suggested that the corona contracts when the inner disk radius keeps small and unchanged. \citet{Buisson2019} reported an increase in QPO frequency when no significant change was found in the inner disc radius. Their findings bring a challenge for models involving geometric effects (orbital or precession) on time-scales related to the inner edge of the disc.
\citet{MA2021} proposed that the type-C QPO comes from a small-scale precessing jet, and the Doppler effect cause the modulation of the observed flux from the jet. 
The X-ray energy decreases along the jet direction, possibly due to the magnetic field and the Compton cooling effect. Considering that the jet twists around the spin axis, the QPOs from the jet base would lag the ones from the jet top if the jet base was observed first. The jet precession model can explain the soft lag observed at high energy ($>$10\,keV) in the hard state. In the jet precession model, the QPO rms (in the LHS) energy dependence above 10\,keV is modulated by the Doppler beaming effect, while the rms amplitude in observed flux depends on the jet speed. The jet speed does not change at different parts of the small-scale jet, thus giving a constant rms in different energy bands. However, the jet precession model was originally built to explain the high-energy LFQPOs, without consideration of the low-energy LFQPOs. Combined with NICER data, we extend the rms energy study to 0.2\,keV. From 0.2\,keV to 2\,keV, the rms increases rapidly with energy, this may indicate the contribution of another component in the lower energy. Here we thus expend the jet precession model by assuming that the observed QPOs are generated by the LT precession of both the jet and the inner disk ring, which are coupled together rigidly, and the jet is launched along the normal direction of the disk ring.  

\begin{figure}
\plotone{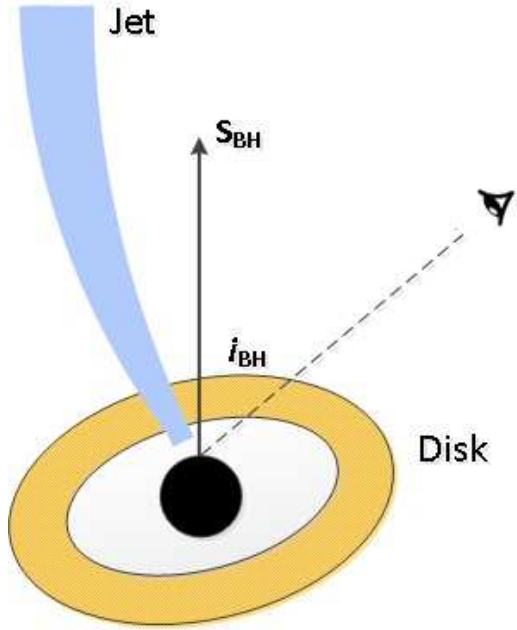}
\caption{The model shows the jet and inner disk ring precess together.
\label{fig:model}}
\end{figure}

The misalignment relative to the BH spin could also cause the accretion disc to precess at a certain rate \citep{Liska2019a, Liska2019b, Schnittman2006}.
\citet{Schnittman2006} developed a precessing ring model, which suggested that the low-energy QPO is generated by a misaligned and precessing inner accretion disk. Such a precessing inner disk would modulate the X-ray emission through relativistic beaming and light-bending effects \citep{Homan2015}. 
\citet{Romero2000} suggested that the inner jet and the innermost part of the accretion disk are coupled in such a way that a precession of the disk induces the precession of the jet in 3C 273, which was suggested to be the case in the binary systems SS433 \citep{Katz1980}. 

Figure \ref{fig:model} gives the geometry of the jet and inner disk ring. When the disk precesses around the BH, it induces the twisted jet precessing around the BH, and the base of the jet and the disk precess together, so they have the same phase of precession. Then the highest energy band ($>$200\,keV, mainly from the lowest part of the jet) has the same sign of phase lag with the lowest energy band ($<1$\,keV, mainly from disk) relative to the energy band of $\sim$10\,keV. 
If the low-energy QPOs are generated from the precessing ring in the accretion disk, then the QPO phase lag and rms in different energy bands may be due to the hybrid contributions from the jet and the accretion disc. Therefore in the hard state, the QPO phase lag increases from $\sim$-0.5 to 0, and the QPO rms increases from 3\% to 10\%.


\begin{figure}
\plotone{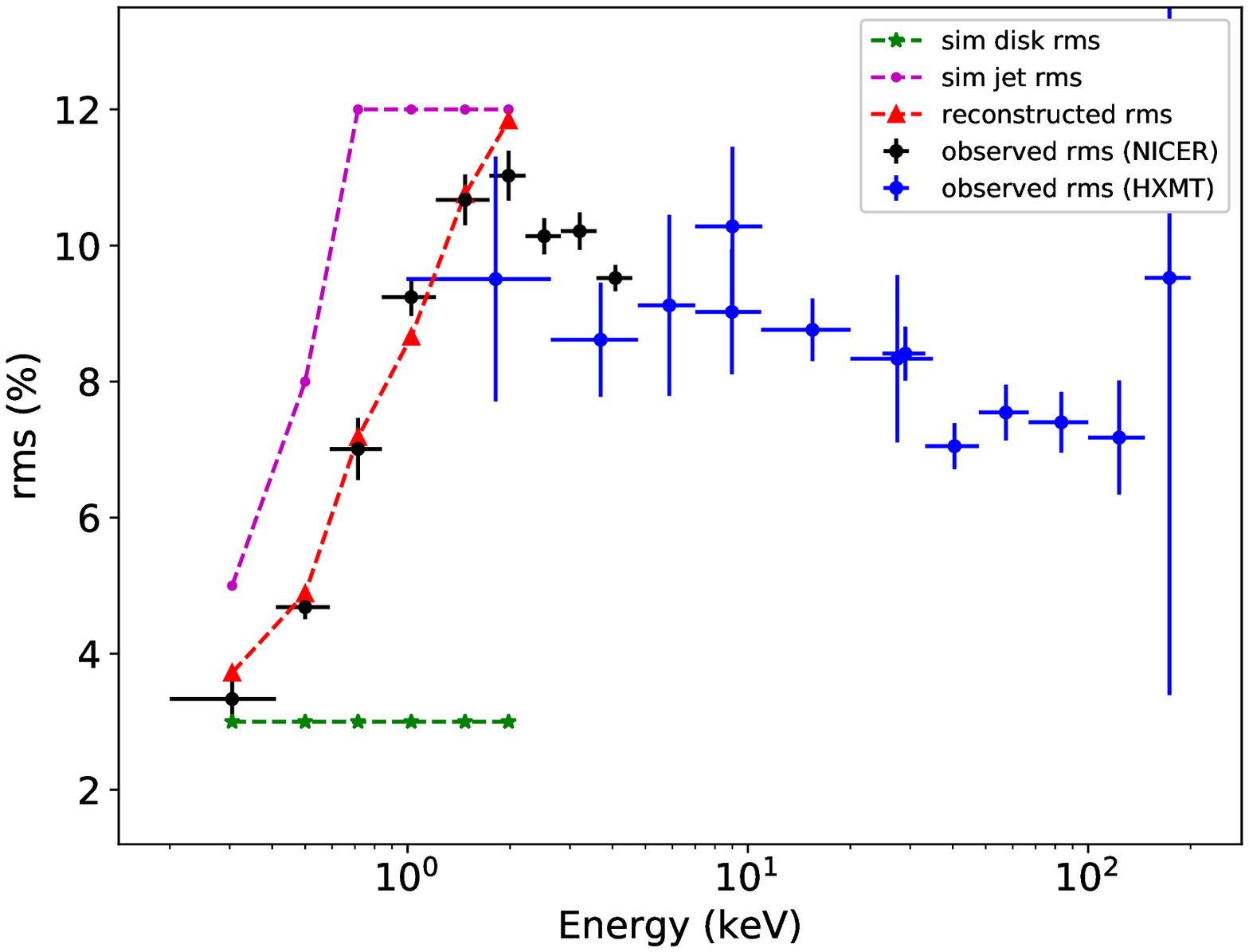}
\plotone{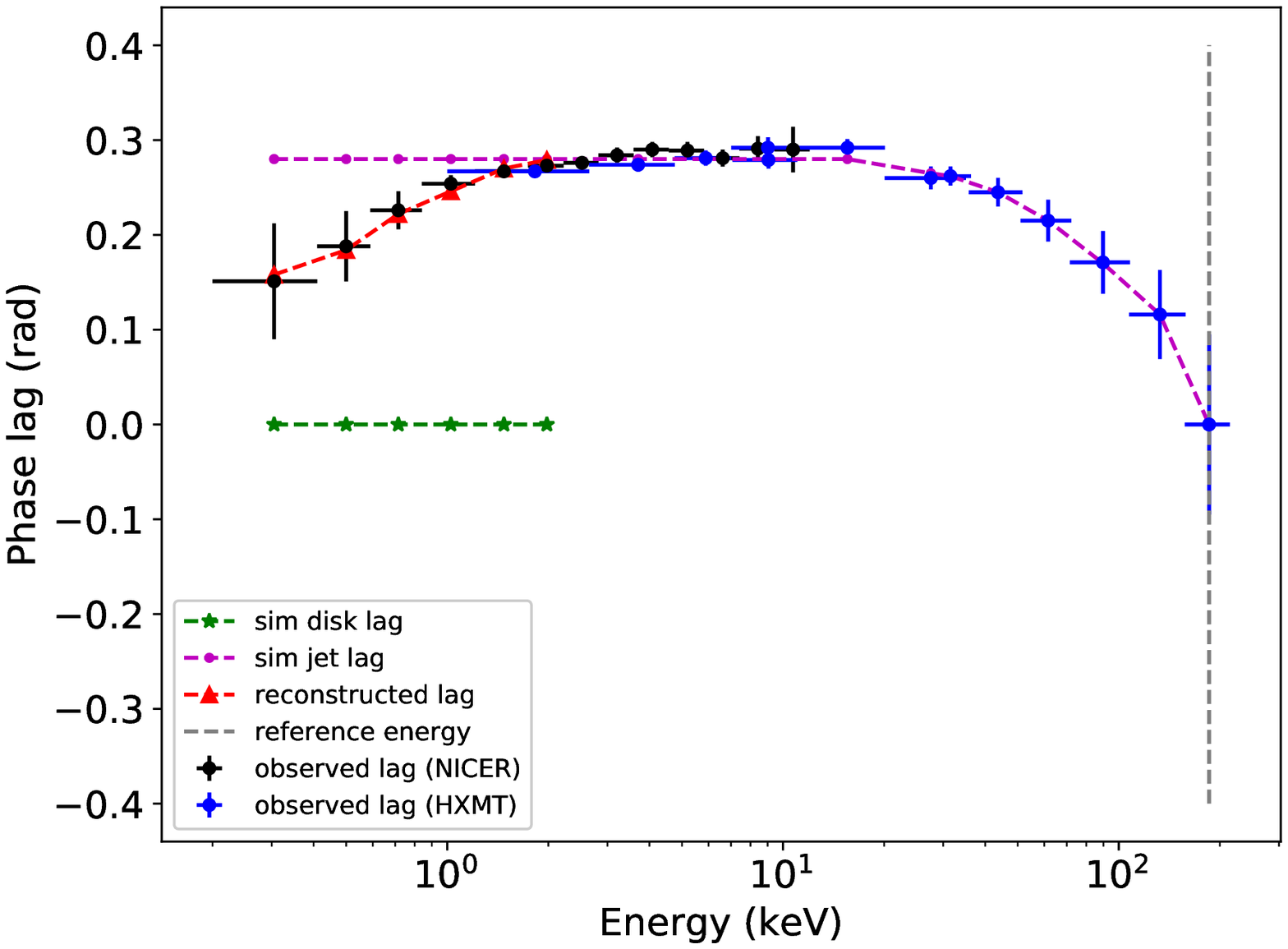}
\caption{Top: The comparison of observed rms and reconstructed rms. Bottom: The comparision of observed phase lags and reconstructed phase lags. The blue dots and black dots represent the observation data of \emph{Insight}-HXMT and NICER, respectively, and the reference energy band of phase lag is the highest energy band which is marked out in grey dashed line. The green stars are the rms and phase lag of disk component with the reference energy band. We assume the disk component precesses with the base of jet, so the phase lags are 0. The purple dots are the rms and phase lags of jet components in the lower energy band. The jet would be collimated at the top of the jet, so the phase lag of lower energy is the same as that at $\sim$10\,keV. The red triangles are the reconstructed rms and phase lag of hybrid light curve. The reconstructed rms and phase lags are consistent with the observed ones.   
\label{fig:simlag}}
\end{figure}
Based on this assumption, we simulate two light curves (jet and disk) and calculate the hybrid light curve phase lag and rms. For simplicity in the toy model, we assume the phase of the jet light curve at the highest energy band $\sim$180\,keV (the base of the jet and also where it is connected to the disk) is 0, and the light curves of the disk in all energy bands have the same phase 0. At a certain Fourier frequency, the signals of each energy band of the jet and disk can be expressed as:
\begin{equation}
    S_{\rm disk} = \sqrt{2}R_{\rm disk} \sin(2\pi ft) + B_{\rm disk},
\end{equation} 
and
\begin{equation}
    S_{\rm jet} = \sqrt{2}R_{\rm jet} \sin(2\pi ft + \phi_{\rm jet}) + B_{\rm jet},
\end{equation}
where $R_{\rm disk}$ and $R_{\rm jet}$ represent the absolute rms of the light curve, which can be calculated with the fractional rms and mean flux $B$. According to the precessing ring model, the QPO amplitude of X-ray light curve fluctuations is a function of the inclination angle of the BH spin axis and the tilt angle of the ring. Assuming the inclination angle $i_{\rm BH}$=63$^{\circ}$ and tilt angle $\Delta \theta$ = 5$^{\circ}$, the rms of the disk is about 3\% \citep{Schnittman2006}. The rms of the jet range from 5\% to 12\% inferred from the multi-wavelength QPO calculation \citep{Jessymol2022}. $B_{\rm disk}$ and $B_{\rm jet}$ are calculated through spectral fitting using {\tt\string tbabs}$\times$({\tt\string diskbb+gaussian+powerlaw)}  following \citet{Wang2020}.  
We combine the light curves of the two components, $S_{\rm disk}$ and $S_{\rm jet}$ at each energy band, and get the hybrid light curve, and then we calculate the phase lag of the combined light curve referenced to the highest energy band. 
Because the disk is bound to the base of the jet, and considering the disk is a rigid body, the phase of the signal of the disk component is the same as the reference phase (i.e., the base of the jet, which is zero). In this way, only the phase of the signal of the jet component at a certain energy is the quantity to be found. The phase lag between the signals at a given energy and the reference energy (180 keV) can be expressed as (considering that the intensity of the jet signal at the reference energy is much greater than that of the disk signal):
\begin{equation}
    \phi(S_{\rm disk\_E} + S_{jet\_E}, S_{jet\_ref} ) = \phi_{\rm obs},
\end{equation} 

Constantly we adjust the signal phase of jet in each energy, so that the left side of equation (3) equals the observed phase lag, and finally obtain the trend of the jet phase with the energy. The results are shown in Figure \ref{fig:simlag} (red triangle line). 
%
%
From the simulation, we can see that our toy model could explain the observed QPO rms and phase lags at different energy bands, assuming the phase of the jet changes with energy. 
%
During the HIMS, the change in QPO phase lag with energy is most likely due to the geometric variability of the disk and jet \citep{Wang2021}, and the detailed geometry of the disk will be presented in another work (Huang et al., in preparation). 




In summary, we have presented the detailed energy-dependent timing analysis of the new BHC MAXI J1820+070 using \emph{Insight}-HXMT and NICER observations in its 2018 outbursts.
For the first time, an energy dependence of the QPO fractional rms, frequency, and phase lags was observed from 0.2\,keV to above 200\,keV. We find that the type-C QPO property could be understood in the frame of Lense-Thirring precession of the jet and the inner disk caused by the black hole's spin.

\begin{acknowledgments}
This work made use of the data from the \emph{Insight}-HXMT mission, a project funded by China National Space Administration (CNSA) and the Chinese Academy of Sciences (CAS). The \emph{Insight}-HXMT team gratefully acknowledges the support from the National Key R\&D Program of China (2021YFA0718500). The authors thank supports from the National Natural Science Foundation of China under Grants U1838111, U1838115, U1838201, U1838202, U2031205, 12122306, 11473027, 11633006, 11673023, 11733009, 12203052. This work was partially supported by International Partnership Program of Chinese Academy of Sciences (Grant No.113111KYSB20190020).

\end{acknowledgments}


\begin{thebibliography}{}
\bibitem[\protect\citeauthoryear{Atri et al.}{2020}]{Atri2020} Atri P., Miller-Jones J.~C.~A., Bahramian A., Plotkin R.~M., Deller A.~T., Jonker P.~G., Maccarone T.~J., et al.\ 2020, \mnras, 493, L81
\bibitem[Alabarta et al.(2022)]{Alabarta2022} Alabarta K., M{\'e}ndez M., Garc{\'\i}a F., Peirano V., Altamirano D., Zhang L., Karpouzas K.\ 2022, \mnras, 514, 2839
\bibitem[\protect\citeauthoryear{Bellavita et al.}{2022}]{Bellavita2022} Bellavita C., Garc{\'\i}a F., M{\'e}ndez M., Karpouzas K., \ 2022, \mnras, 515, 2099
\bibitem[Belloni, Psaltis \& van der Klis(2002)]{Belloni2002} Belloni, T., Psaltis, D., \& van der Klis, M.\ 2002, \apj, 572, 392
\bibitem[Belloni et al.(2005)]{Belloni2005} Belloni, T., Homan, J., Casella, P.,  van der Klis M., Nespoli E., Lewin W.~H.~G., Miller J.~M., M{\'e}ndez M.,\ 2005, \aap, 440, 207
\bibitem[Belloni(2010)]{Belloni2010} Belloni, T.\ 2010, Lecture Notes in Physics, Berlin Springer Verlag, 794
\bibitem[Belloni et al.(2020)]{Belloni2020} Belloni T.~M., Zhang L., Kylafis N. D., Reig P. \& Altamirano D.\ 2020, \mnras, 496, 4366
\bibitem[Bright, Fender \& Motta(2018)]{Bright2018} Bright, J., Fender, R., \& Motta, S.\ 2018, ATel, 11420
\bibitem[Buisson et al.(2018)]{Buisson2018} Buisson, D., Fabian, A., Alston, W., et al.\ 2018, ATel, 11578
\bibitem[Buisson et al.(2019)]{Buisson2019} Buisson, D., Fabian, A., Barret, D., et al.\ 2019, \mnras, 490, 1350
\bibitem[Cabanac et al.(2010)]{Cabanac2010} Cabanac, C., Henri, G., Petrucci, P.-O., et al.\ 2010, \mnras, 404, 738

\bibitem[Casella et al.(2004)]{Casella2004} Casella, P., Belloni, T., Homan, J., \& Stella, L.\ 2004, \aap, 426, 587
\bibitem[Casella, Belloni \& Stella(2005)]{Casella2005} Casella, P., Belloni, T., \& Stella, L.\ 2005, \apj, 629, 403
\bibitem[Denisenko(2018)]{Denisenko2018} Denisenko, D.\ 2018, ATel, 11400
\bibitem[\protect\citeauthoryear{Garg, Misra, \& Sen}{2022}]{2022MNRAS.514.3285G} Garg A., Misra R., Sen S., 2022, MNRAS, 514, 3285 
\bibitem[\protect\citeauthoryear{Garc{\'\i}a et al.}{2021}]{Garcia2021} Garc{\'\i}a F., M{\'e}ndez M., Karpouzas K., Belloni T., Zhang L., Altamirano D.,\ 2021, MNRAS, 501, 3173 
\bibitem[Gierli\'{n}ski \& Zdziarski(2005)]{Gierlinski2005} Gierli\'{n}ski, M., \& Zdziarski, A.~A.\ 2005, \mnras, 363, 1349
\bibitem[\protect\citeauthoryear{Guan et al.}{2021}]{Guan2021} Guan J., Tao L., Qu J.~L., Zhang S.~N., Zhang W., Zhang S., Ma R.~C., et al.\ 2021, \mnras, 504, 2168
\bibitem[\protect\citeauthoryear{Guo et al.}{2020}]{Guo2020} Guo C.-C., Liao J.-Y., Zhang S., Zhang J., Tan Y., Song L.-M., Lu F.-J., et al.\ 2020, JHEAp, 27, 44
\bibitem[Homan et al.(2015)]{Homan2015} Homan, J., Fridriksson, J. K.,  \& Remillard, R. A.\ 2015, \apj, 812, 80
\bibitem[Homan et al.(2018a)]{Homan2018a} Homan, J., Altamirano, D., Arzoumanian, Z., et al.\ 2018a, ATel, 11576
\bibitem[Homan et al.(2018b)]{Homan2018b} Homan, J., Uttley, P., Gendreau, Z., et al.\ 2018b, ATel, 11820
\bibitem[Homan et al.(2018c)]{Homan2018c} Homan, J., Uttley, P., Gendreau, Z., et al.\ 2018c, ATel, 11823
\bibitem[Homan et al.(2018d)]{Homan2018d} Homan, J., Stevens, A.~L., Altamirano, D., et al.\ 2018d, ATel, 12068
\bibitem[Homan et al.(2020)]{Homan2020} Homan, J., Bright, J., Motta, S., et al.\ 2020, \apjl, 891, L29
\bibitem[Huang et al.(2018)]{Huang2018} Huang, Y., Qu, J. L., Zhang, S.~N., et al.\ 2018, \apj, 866, 122
\bibitem[\protect\citeauthoryear{Huppenkothen et al.}{2019}]{Huppenkothen2019} Huppenkothen D., Bachetti M., Stevens A.~L., Migliari S., Balm P., Hammad O., Khan U.~M., et al., 2019, ApJ, 881, 39
\bibitem[Ingram, Done \& Fragile(2009)]{Ingram2009} Ingram, A., Done, C., \& Fragile, P.~C.\ 2009, \mnras, 397, L101
\bibitem[Ingram et al.(2016)]{Ingram2016} Ingram, A., van der Klis, M., Middleton, M., et al.\ 2016, \mnras, 461, 1967
\bibitem[Ingram(2020)]{Ingram2020} Ingram, A., Motta, S. \ 2020, New Astronomy Reviews, arXiv:2001.08758
\bibitem[\protect\citeauthoryear{Karpouzas et al.}{2020}]{Karpouzas2020} Karpouzas K., M{\'e}ndez M., Ribeiro E.~M., Altamirano D., Blaes O., Garc{\'\i}a F.,\ 2020, \mnras, 492, 1399 
\bibitem[\protect\citeauthoryear{Karpouzas et al.}{2021}]{Karpouzas2021} Karpouzas K., M{\'e}ndez M., Garc{\'\i}a F., Zhang L., Altamirano D., Belloni T., Zhang Y.,\ 2021, \mnras, 503, 5522
\bibitem[\protect\citeauthoryear{Katz}{1980}]{Katz1980} Katz J.~I.\  1980, \apjl, 236, L127.
\bibitem[Kara et al.(2019)]{Kara2019} Kara E. et al., 2019, Nature, 565, 198
\bibitem[Kawamuro et al.(2018)]{Kawamuro2018} Kawamuro, T., Negoro, H., Yoneyama, T., et al.\ 2018, ATel, 11399
\bibitem[Kennea \& Siegel(2018)]{Kennea2018} Kennea, J. A., \& Siegel, M. H.\ 2018, ATel, 11404
 
\bibitem[Liao et al.(2020a)]{Liao2020a} Liao J.-Y., Zhang S., Lu X.-F., Zhang J., Li G., Chang Z., Chen Y.-P., et al.\ 2020a, JHEAp, 27, 14
\bibitem[Liao et al.(2020b)]{Liao2020b} Liao J.-Y., Zhang S., Chen Y., Zhang J., Jin J., Chang Z., Chen Y.-P., et al.\ 2020b, JHEAp, 27, 24
\bibitem[Li et al.(2013)]{Li2013a} Li, Z.~B., Qu, J.~L., Song, L. M., et al.\ 2013a, \mnras, 428, 1704
\bibitem[Li et al.(2013)]{Li2013b} Li, Z.~B., Zhang, S., Qu, J.~L., et al.\ 2013b, \mnras, 433, 412
\bibitem[\protect\citeauthoryear{Liska et al.}{2019a}]{Liska2019a} Liska M., Hesp C., Tchekhovskoy A., Ingram A., van der Klis M., \& Markoff S.~B.\ 2019, arXiv:1901.05970
\bibitem[\protect\citeauthoryear{Liska et al.}{2019b}]{Liska2019b} Liska M., Tchekhovskoy A., Ingram A., \& van der Klis M.\ 2019, \mnras, 487, 550
\bibitem[\protect\citeauthoryear{Liu et al.}{2021}]{Liu2021} Liu H.-X., Huang Y., Xiao G.-C., Bu Q.-C., Qu J.-L., Zhang S., Zhang S.-N., et al.\ 2021, RAA, 21, 070 
\bibitem[Ma et al.(2021)]{MA2021} Ma, X., Tao, L., Zhang, S.~N., et al.\ 2021, Nat.Astron, 5,94
\bibitem[\protect\citeauthoryear{M{\'e}ndez \& van der Klis}{1997}]{Mendez1997} M{\'e}ndez M., van der Klis M.,\ 1997, \apj, 479, 926
\bibitem[\protect\citeauthoryear{M{\'e}ndez et al.}{2022}]{Mendez2022} M{\'e}ndez M., Karpouzas K., Garc{\'\i}a F., Zhang L., Zhang Y., Belloni T.~M., Altamirano D.,\ 2022, Nat.Astron, 6, 761
\bibitem[Mereminskiy et al.(2018)]{Mereminskiy2018} Mereminskiy, I. A., Grebenev, S. A., Molkov, S. V., et al.\ 2018, ATel, 11488
\bibitem[Miyamoto et al.(1991)]{Miyamoto1991} Miyamoto, S., Kimura, K., Kitamoto, S., Dotani, T., Ebisawa, K.\ 1991, \apj, 383, 784
\bibitem[Molteni, Sponholz, \& Chakrabarti(1996)]{Molteni1996} Molteni D., Sponholz H., Chakrabarti S.~K.\ 1996, \apj, 457, 805. doi:10.1086/176775
\bibitem[Motta et al.(2015)]{Motta2015} Motta, S., Casella, P., Henze, M., et al.\ 2015, \mnras, 447, 2059
\bibitem[Motta(2016)]{Motta2016} Motta, S. E.\ 2016, AN, 337, 398
\bibitem[Nowak et al.(1999)]{Nowak1999} Nowak, M.~A., Vaughan, B.~A., Wilms, J., et al.\ 1999, \apj, 510, 874
\bibitem[Qu et al.(2010)]{Qu2010} Qu, J.~L., Lu, F.~J., Lu, Y., et al.\ 2010, \apj, 710, 836
\bibitem[Rodriguez et al.(2004)]{Rodriguez2004} Rodriguez, J., Corbel, S., Kalemci, E., et al.\ 2004, \apj, 612, 1018
\bibitem[\protect\citeauthoryear{Romero et al.}{2000}]{Romero2000} Romero G.~E., Chajet L., Abraham Z., Fan J.~H.,\ 2000, A\&A, 360, 57
\bibitem[Schnittman, Homan \& Miller(2006)]{Schnittman2006} Schnittman, J.~D., Homan, J., \& Miller, J.~M.\ 2006, \apj, 642, 420
\bibitem[\protect\citeauthoryear{Steiner et al.}{2010}]{Steiner2010} Steiner J.~F., McClintock J.~E., Remillard R.~A., Gou L., Yamada S., Narayan R., 2010, ApJL, 718, L117
\bibitem[Stevens \& Uttley(2016)]{Stevens2016} Stevens, A. L., \& Uttley, P.\ 2016, \mnras, 460, 2796
\bibitem[Stiele et al.(2020)]{Stiele2020} Stiele, H., \& Kong, A. K. H.\ 2020, \apj, 889, 142S
\bibitem[Tagger \& Pellat(1999)]{Tagger1999} Tagger, M., \& Pellat, R.\ 1999, \aap, 349, 1003
\bibitem[\protect\citeauthoryear{Thomas et al.}{2022}]{Jessymol2022} Thomas J.~K., Buckley D.~A.~H., Charles P.~A., Paice J.~A., Potter S.~B., Steiner J.~F., Lasota J.-P., et al.\ 2022, \mnras, 513, L35
\bibitem[Torres et al.(2019)]{Torres2019} Torres, M., Casares, J., et al.\ 2019, \apjl, 882, L21
\bibitem[\protect\citeauthoryear{Torres et al.}{2020}]{Torres2020} Torres M.~A.~P., Casares J., Jim{\'e}nez-Ibarra F., {\'A}lvarez-Hern{\'a}ndez A., Mu{\~n}oz-Darias T., Armas Padilla M., Jonker P.~G., et al.\ 2020, \apjl, 893, L37. 
\bibitem[van der Klis et al.(2006)]{Klis2006} Van der Klis, M.\ 2006, in Compact Stellar X-ray Sources, ed. W. Lewin \& M. van der Klis (Cambridge: Cambridge Univ. Press), 39
\bibitem[van den Eijnden, Ingram \& Uttley(2016)]{Eijnden2016} van den Eijnden, J., Ingram, A., Uttley, P.\ 2016, \mnras, 458, 3655
\bibitem[Van den Eijnden et al.(2017)]{VandenEijnden2017} van den Eijnden, J., Ingram, A., Uttley, P., et al.\ 2017, \mnras, 464, 2643
\bibitem[Vaughan \& Nowak(1997)]{Vaughan1997} Vaughan, B.~A., \& Nowak, M.~A.\ 1997, \apjl, 474, L43
\bibitem[Wang et al.(2021)]{Wang2021} Wang J., Mastroserio G., Kara E., Garc{\'\i}a J.~A., Ingram A., Connors R., van der Klis M., et al. \ 2021, \apjl, 910, L3.
\bibitem[\protect\citeauthoryear{Wang et al.}{2020}]{Wang2020} Wang Y., Ji L., Zhang S.~N., M{\'e}ndez M., Qu J.~L., Maggi P., Ge M.~Y., et al., 2020, ApJ, 896, 33. doi:10.3847/1538-4357/ab8db4
\bibitem[Yang et al.(2022)]{Yang2022} Yang Z.-X., Zhang L., Bu Q.-C., Huang Y., Liu H.-X., Yu W., Wang P.~J., et al.\ 2022, \apj, 932, 7.
\bibitem[Yadav et al.(2016)]{Yadav2016} Yadav, J. S., Misra, R., Verdhan Chauhan, J., et al.\ 2016, \apj, 833, 27
\bibitem[You et al.(2018)]{You2018} You, B., Bursa, M., \& \'{Z}ycki, P. T. \ 2018, \apj, 858, 82
\bibitem[\protect\citeauthoryear{You et al.}{2021}]{YOU2021} You B., Tuo Y., Li C., Wang W., Zhang S.-N., Zhang S., Ge M., et al.\ 2021, NatCo, 12, 1025. 
\bibitem[Yu et al.(2018a)]{Yu2018a} Yu, Wenfei., Zhang, Jujia., Yan, Zhen., Wang, Xiaofeng., Bai, Jinming.\ 2018, ATel, 11510
\bibitem[Yu et al.(2018b)]{Yu2018b} Yu, Wenfei., Lin, Jie., Mao, Dongming., Zhang, Jujia., Yan, Zhen., Bai, Jinming.\ 2018, ATel, 11591
\bibitem[Zampieri et al.(2018)]{Zampieri2018} Zampieri. L, Fiori, M., Burtovoi, A., et al.\ 2018, ATel, 11723
\bibitem[Zdziarski(2003)]{Zdziarski2003} Zdziarski, A.~A.\ 2003, Bulletin of the American Astronomical Society, 35, 29.02
\bibitem[Zhang et al.(2017)]{Zhang2017} Zhang, L., Wang, Yanan., M\'{e}ndez, M., et al.\ 2017, \apj, 845, 143
\bibitem[Zhang et al.(2020)]{zhang2020}Zhang, S.~N., Li, T.~P., Lu, F.~J., et al.\ 2020, Sci. China-Phys. Mech. Astron., 63, 249502
\bibitem[\protect\citeauthoryear{Zhang et al.}{2022}]{zhang2022} Zhang Y., M{\'e}ndez M., Garc{\'\i}a F., Zhang S.-N., Karpouzas K., Altamirano D., Belloni T.~M., et al.,\ 2022, MNRAS, 512, 2686
\bibitem[\protect\citeauthoryear{Zhao et al.}{2021}]{zhao2021} Zhao X., Gou L., Dong Y., Tuo Y., Liao Z., Li Y., Jia N., et al.\ 2021, \apj, 916, 108 
\bibitem[\protect\citeauthoryear{Zhou et al.}{2022}]{Zhou2022} Zhou D.-K., Zhang S.-N., Song L.-M., Qu J.-L., Zhang L., Ma X., Tuo Y.-L., et al.\ 2022, \mnras, 515, 1914

\end{thebibliography}
\end{document}